\documentclass{aa}
\usepackage{natbib}
\bibpunct{(}{)}{;}{a}{}{,} 

\usepackage{graphicx}
\usepackage[varg]{txfonts}
\def\teff{T_{\rm eff}}
\def\aml{\alpha_{\rm{MLT}}}

\def\msun{\rm{M}_\sun}

\def\rsun{\rm{R}_\sun}

\begin{document}

\title{The Aarhus Red Giants Challenge I}
\subtitle{Stellar structures in the red giant branch phase}
\author{
V.~Silva Aguirre\inst{\ref{instsac}} \and J.~Christensen-Dalsgaard\inst{\ref{instsac}} \and S.~Cassisi\inst{\ref{instinaf},\ref{instinaf2}} \and M.~Miller~Bertolami\inst{\ref{instLP1},\ref{instLP2},\ref{instmpa}} \and A.~Serenelli\inst{\ref{instald1},\ref{instald2}} \and D.~Stello\inst{\ref{instsyd2},\ref{instsyd}, \ref{instsac}} \and A.~Weiss\inst{\ref{instmpa}} \and
G.~Angelou\inst{\ref{instmpa},\ref{instmps}} \and C.~Jiang\inst{\ref{instchi}} \and Y.~Lebreton\inst{\ref{instleb1},\ref{instleb2}} \and F.~Spada\inst{\ref{instmps}} \and
E.~P.~Bellinger\inst{\ref{instsac},\ref{instmps}} \and S.~Deheuvels\inst{\ref{instdeh}} \and R.~M.~Ouazzani\inst{\ref{instleb1},\ref{instsac}} \and 
A.~Pietrinferni\inst{\ref{instinaf}} \and  J.~R.~Mosumgaard\inst{\ref{instsac}} \and R.~H.~D.~Townsend\inst{\ref{instrich}} \and
T.~Battich\inst{\ref{instLP1},\ref{instLP2}} \and D.~Bossini\inst{\ref{instpor},\ref{instbir},\ref{instsac}} \and T. Constantino\inst{\ref{instexe}} \and P.~Eggenberger\inst{\ref{instegg}} \and S.~Hekker\inst{\ref{instmps},\ref{instsac}} \and A.~Mazumdar\inst{\ref{instanw}} \and A.~Miglio\inst{\ref{instbir},\ref{instsac}} \and K.~B.~Nielsen\inst{\ref{instsac}} \and M.~Salaris\inst{\ref{instsal}}
}
\institute{
Stellar Astrophysics Centre, Department of Physics and Astronomy, Aarhus University, Ny Munkegade 120, DK-8000 Aarhus C, Denmark\label{instsac} \and
INAF-Astronomical Observatory of Abruzzo, Via M. Maggini sn, I-64100 Teramo, Italy\label{instinaf} \and
INFN - Sezione di Pisa, Largo Pontecorvo 3, 56127 Pisa, Italy\label{instinaf2} \and
Instituto de Astrof\'isica de La Plata, UNLP-CONICET, La Plata, Paseo del Bosque s/n, B1900FWA, Argentina\label{instLP1}\and
Facultad de Ciencias Astron\'omicas y Geof\'isicas, UNLP, La Plata, Paseo del Bosque s/n, B1900FWA, Argentina\label{instLP2}\and
Max-Planck-Institut f\"{u}r Astrophysics, Karl Schwarzschild Strasse 1, 85748, Garching, Germany\label{instmpa} \and
Instituto de Ciencias del Espacio (ICE, CSIC), Campus UAB, Carrer de Can Magrans, s/n, 08193 Cerdanyola del Valles, Spain\label{instald1} \and
Institut d'Estudis Espacials de Catalunya (IEEC), Gran Capita 4, E-08034, Barcelona, Spain\label{instald2} \and
School of Physics, University of New South Wales, NSW, 2052, Australia\label{instsyd2} \and
Sydney Institute for Astronomy, School of Physics, University of Sydney, NSW 2006, Australia\label{instsyd} \and
Max-Planck-Institut f\"{u}r Sonnensystemforschung, Justus-von-Liebig-Weg 3, 37077, G\"{o}ttingen, Germany\label{instmps} \and
School of Physics and Astronomy, Sun Yat-Sen University, Guangzhou, 510275, China \label{instchi} \and
LESIA, Observatoire de Paris, PSL Research University, CNRS, Sorbonne Universit\'e, Univ. Paris Diderot, Sorbonne Paris Cit\'e, Meudon 92195, France\label{instleb1} \and
Univ Rennes, CNRS, IPR (Institut de Physique de Rennes) - UMR 6251, F-35000 Rennes, France\label{instleb2} \and
IRAP, Universit\'e de Toulouse, CNRS, CNES, UPS, Toulouse, France\label{instdeh} \and
Department of Astronomy, 2535 Sterling Hall 475 N. Charter Street, Madison, WI 53706-1582, USA\label{instrich} \and
Instituto de Astrof\'isica e Ci\^encias do Espa\c co, Universidade do Porto, CAUP, Rua das Estrelas, PT-4150-762 Porto, Portugal\label{instpor} \and
School of Physics and Astronomy, University of Birmingham, Birmingham, B15 2TT, UK\label{instbir} \and
Physics and Astronomy, University of Exeter, Exeter, EX4 4QL, United Kingdom\label{instexe} \and
Observatoire de Gen\`eve, Universit\'e de Gen\`eve, 51 Ch. des Maillettes, CH-1290 Sauverny, Suisse\label{instegg} \and
Homi Bhabha Centre for Science Education, TIFR, V. N. Purav Marg, Mankhurd, Mumbai 400088, India\label{instanw} \and
Astrophysics Research Institute, Liverpool John Moores University, 146 Brownlow Hill, Liverpool L3 5RF, UK\label{instsal}
}
\abstract
{With the advent of space-based asteroseismology, determining accurate properties of red-giant stars using their observed oscillations has become the focus of many investigations due to their implications in a variety of fields in astrophysics. Stellar models are fundamental in predicting quantities such as stellar age, and their reliability depends critically on the numerical implementation of the physics at play in this evolutionary phase.}
{We introduce the {\it Aarhus Red Giants Challenge}, a series of detailed comparisons between widely used stellar evolution and oscillation codes aiming at establishing the minimum level of uncertainties in properties of red giants arising solely from numerical implementations. We present the first set of results focusing on stellar evolution tracks and structures in the red-giant-branch phase (RGB).}
{Using 9 state-of-the-art stellar evolution codes, we defined a set of input physics and physical constants for our calculations and calibrated the convective efficiency to a specific point on the main sequence. We produced evolutionary tracks and stellar structure models at fixed radius along the red-giant branch for masses of $1.0~\msun$, $1.5~\msun$, $2.0~\msun$, and $2.5~\msun$, and compared the predicted stellar properties.}
{Once models have been calibrated on the main sequence we find a residual spread in the predicted effective temperatures across all codes of $\sim20$~K at solar radius and $\sim$30-40~K in the RGB regardless of the considered stellar mass. The predicted ages show variations of 2-5\% (increasing with stellar mass) which we track down to differences in the numerical implementation of energy generation. The luminosity of the RGB-bump shows a spread of about 10\% for the considered codes, which translates into magnitude differences of $\sim0.1$~mag in the optical V-band. We also compare the predicted [C/N] abundance ratio and found a spread of 0.1~dex or more for all considered masses.}
{Our comparisons show that differences at the level of a few percent still remain in evolutionary calculations of red giants branch stars despite the use of the same input physics. These are mostly due to differences in the energy generation routines and interpolation across opacities, and call for further investigations on these matters in the context of using properties of red giants as benchmarks for astrophysical studies. We make available all our evolutionary calculations and models through the website\thanks{\url{https://github.com/vsilvagui/aarhus_RG_challenge}} of the workshops.}
\maketitle
\titlerunning{The Aarhus red giants challenge I}
\authorrunning{Silva Aguirre et al.}
\section{Introduction}\label{sec_int}
Red giants of low and intermediate mass are cool luminous stars found in three evolutionary phases: on the red-giant branch (RGB), during core-helium burning (or clump), and on the asymptotic giant branch (AGB). They are of key importance in many fields of astrophysics, for example as benchmarks for testing stellar evolution theory in star clusters. Thanks to their high intrinsic luminosity, red giants are perfectly suited to explore distant regions of the Galaxy where accurate observations of fainter stars become challenging.

Determining precise stellar properties of field giants is extremely difficult when using traditional techniques such as matching effective temperature, gravity, and composition to stellar tracks or isochrones. The reason is that giant stars in different evolutionary stages and spanning a wide range in mass overlap in the observational plane (e.g., in the  Kiel or Color-Magnitude diagram) well within the observational errors typically obtained with e.g., spectroscopy or photometry. The net result is that properties of red giants determined in this manner are largely dominated by the statistical uncertainties \citep[see e.g.,][]{Serenelli:2013fz, SilvaAguirre:2016km}.

The study of red-giant stars has gone through a revolution with the advent of asteroseismic data from space missions. The {\it CoRoT} \citep{Baglin:2006ui} and {\it Kepler} \citep{Gilliland:2010bb} satellites have inspired a new paradigm of precisely determined stellar properties. By measuring the brightness variations in a large number of red giants across the Galaxy, these missions have provided new insights into the otherwise inaccessible deep interior of stars. Among the most striking results from asteroseismology are the detection of non-radial pulsation modes in red giant stars \citep{DeRidder:2009cd}. These can help distinguishing between RGB and clump stars \citep{Bedding:2011il}, they can provide a measurement of the rotation profile from the inner core to the envelope \citep[e.g.,][]{Beck:2012en,Mosser:2012dj,Cantiello:2014dh}, of the efficiency of core mixing during the helium-burning phase \citep[e.g.,][]{Montalban:2013gf,Constantino:2015fu,Bossini:2017cn}, and the possible prevalence of fossil magnetic fields in their cores \citep{Stello:2016jg,2017A&A...598A..62M}.

The availability of asteroseismic data for thousands of red giants also allows the study of ensembles of stars, advancing into the field of Galactic archaeology \citep[e.g.,][]{Miglio:2013hh,2018MNRAS.475.5487S}. Properties based on asteroseismic data now include distances \citep{SilvaAguirre:2012du,Rodrigues:2014it}, masses and radii \citep{Casagrande:2014bd,Pinsonneault:2015kd,2016ApJ...822...15S}, and most recently ages for RGB and clump stars \citep[][]{2016MNRAS.455..987C,2017A&A...597A..30A,2018ApJS..239...32P}. All these properties are determined to unprecedented level of precision by combining the asteroseismic information with stellar evolution and pulsation calculations. If not only precise but also accurate, asteroseismically derived stellar properties have the potential to serve as a benchmark to improve our understanding of stellar structure and evolution as well as the processes that have been shaping our Milky Way into what it is today.

Inspired by the high-quality asteroseismic data obtained by the CoRoT and {\it Kepler} missions, and its clear potential for furthering scientific understanding in different fields of astrophysics, there is a fast growing body of literature devoted to validation of masses, radii, distances and ages determined from asteroseismology. This work includes comparing the seismic-inferred properties to those obtained empirically using interferometry \citep{Huber:2012iv,White:2013bu}, parallaxes \citep{DeRidder:2016bl,Huber:2017fg,2018MNRAS.476.1931S}, binary stars \citep[e.g.,][]{Frandsen:2013ih,2016ApJ...832..121G,2018MNRAS.478.4669T}, and open clusters \citep{Brogaard:2011jx,Basu:2011cc,2018MNRAS.476.3729B}, to name just a few. Still, our ability to accurately determine asteroseismically-inferred stellar properties ultimately relies on having a realistic theoretical description of the structure and evolution of stars that can reproduce the features given by classical and seismic observations.

Many large compilations of theoretical tracks and isochrones produced by different groups are freely available and widely used by the community to determine stellar properties of red giants \citep[e.g.,][]{Pietrinferni:2004im,Dotter:2008ga,Bressan:2012bx,2016ApJ...823..102C,Spada:2017ev,Hidalgo:2018dy}. These sets are computed assuming a certain combination of micro-and macro-physics and by calibrating the convective efficiency to match the properties of the Sun, ensuring that the 1~M$_\sun$ track has the correct solar temperature, luminosity, and surface composition at the solar age. Beyond this single calibration point, tracks computed by different groups predict different properties for the Sun's subsequent evolution. These differences are inherent to the code's numerical schemes (i.e., equation solvers, interpolations over tables, etc.) and differ from those arising by variations in the input physics. However, when comparing compilations of tracks and isochrones it is challenging to determine which differences belong to the chosen micro-and macro-physics and which would remain if the considered physical description was exactly the same. The latter are of utmost importance, as they define the maximum precision that can be attained when determining stellar properties based on evolutionary calculations, regardless of the nature of the constraints applied (i.e., spectroscopy, interferometry, or asteroseismology).

With this in mind, we started a series of workshops known as the {\it Aarhus Red Giants Challenge}. The aim of these meetings was to gather experts working directly in the development of evolutionary and pulsation codes, produce sets of benchmark tracks and models with clearly defined input physics, compare the results from these different implementations, find and understand the origin of any discrepancies, and quantify the intrinsic uncertainties when determining stellar properties of red giants arising solely from the numerical methods applied by each code. Based on this information, the ultimate goal of these workshops is to define the best asteroseismic diagnosis methods for red giant stars and agree on a minimum set of observables necessary to characterise them to various levels of precision. This is in a sense an exercise focused on red giant stars along the same lines as what was performed in the ESTA framework for main-sequence models \citep{Lebreton:2008kt}.

Here we present the first set of calibration and science cases of red-giant-branch models produced for the {\it Aarhus Red Giants Challenge}. Besides a general description of all models produced, in this paper we focus on the structural and evolutionary differences produced by the participating stellar evolution codes with particular emphasis on (1) directly observed features (such as differences in effective temperature, chemical abundance, or the RGB-bump luminosity), (2) parameters that affect derived quantities such as age (convective-core size and energy generation rates), and (3) those that have an impact in asteroseismic observations (such as the interior hydrogen profile). Detailed comparisons of asteroseismic quantities and oscillation frequencies for the science cases presented here are the subject of the accompanying paper (Christensen-Dalsgaard et al. 2019, hereafter Paper~II), while the analysis of structures in the clump phase and studies of the impact of changes in the input physics will be the subject of subsequent publications (Miller~Bertolami et al., in preparation; Angelou et al., in preparation). A full account of the activities carried out as part of this challenge is available on the website of the workshops and includes a description of the science cases and all models and tracks computed.
\section{Input physics and physical constants}\label{sec_inpphys}
We started our comparisons by defining a common set of input physics to be used in all exercises. We considered the original NACRE compilation of nuclear reactions \citep{Angulo:1999kp} without any updates to the rates, the OPAL opacities and the 2005 version of the equation of state \citep{Iglesias:1996dp, 2002ApJ...576.1064R}, the Potekhin conductive opacities \citep{Cassisi:2007ey} and the \citet{Grevesse:1993vd} mixture of solar abundances. A simple Eddington relation has been used for the atmospheric stratification and the evolution has been started from the Zero Age Main Sequence (ZAMS). Convection has been treated under the mixing-length theory (MLT), with an efficiency calibrated using the solar radius as defined in Section~\ref{sec_suncal}. The boundary of convective regions is defined using the Schwarszchild criterion \citep{Schwarzschild:1958jy}, and we have not included the effects of overshooting, microscopic diffusion, nor mass loss in this set of calculations. The energy lost by neutrinos in nuclear reactions is taken into account as a decrease of the net energy {\it Q} released in each reaction. Although some of these ingredients are not completely up-to-date, they provide a common framework that all evolutionary codes participating in the challenge are currently able to include.

When comparing the internal structures we defined a maximum acceptable convergence for a model of $\alpha$ solar masses at $\beta$ solar radii as follows:
\begin{equation}
\centering
\Delta_\mathrm{convergence} = \left|1.0 - \frac{G_\mathrm{\tt code} M_\mathrm{\tt code}/{R_\mathrm{\tt code}}^3}{G (\alpha \times \mathrm{M}_\odot)/{(\beta \times \mathrm{R}_\odot)^{3}}} \right| \le 2 \times 10^{-4}\,,
\label{eq:converg}
\end{equation}
\noindent{where $G_{\tt code}$ is the gravitational constant assumed by each evolutionary code, and $M_{\tt code}$, $R_{\tt code}$ are the mass and radius of the calculated model}. We emphasize here the role of the gravitational constant $G$, which is only known to a relative precision of $10^{-5}$ \citep{2016RvMP...88c5009M}. There are differences in the gravitational constant adopted by each evolutionary code included in this challenge, and to further reduce potential sources of discrepancies we have defined a set of physical constants adopted in our calculations which we summarize in Table~\ref{tab:physconst}.

The convergence criterion is chosen to scale with the mean stellar density due to the dependence of the adiabatic oscillation frequencies on this quantity \citep[cf.][]{2010aste.book.....A}. By enforcing this criterion we ensure that differences in the asymptotic large frequency separation are not a consequence of differences in the mass and radius of the models but only on the details of the adiabatic sound speed profile \citep[and therefore in the treatment of opacities, equation of state, etc. See e.g.,][]{2013ASPC..479...61B}. The tolerance has been set to $2\times 10^{-4}$ as a compromise between ease of finding the required model across an evolutionary track and reproducing the acoustic modes of oscillation at a level better than the current uncertainties of the longest seismic observations from the {\it Kepler} mission \citep[below the $\sim$0.1~$\mu$Hz level, see e.g.,][]{2016MNRAS.456.2183D,Lund:2017ez,2018ApJS..236...42Y}. We note that in most cases the comparison points were chosen at specific radii for a given mass value, which defines the quantities $\alpha$ and $\beta$ in the denominator of Eq.~\ref{eq:converg}. Since the gravitational constant adopted by all codes is the same, once the quantities $\alpha$ and $\beta$ are defined for each science case the convergence criterion depends only on a ratio between mass and radius (in this case the mean density) of the evolutionary model in question. Because some codes consider the mass change during the evolution produced by the nuclear energy release, $M_\mathrm{model}$ is not exactly the same as $\alpha\,\times\,$M$_\odot$ when the model reaches the radius $\beta\,\times\,$R$_\odot$. These differences are compensated with variations in radius to reach the required precision and therefore ensure a consistent mean stellar density across all models compared at a given radius. Our convergence criterion is nevertheless generally applicable to cases in which different physical constants or input physics are used that can vary the quantities entering Eq.~\ref{eq:converg}, and the results of that comparison is the subject of the study by Angelou et al. (in preparation).
\begin{table}
\caption{Adopted physical constants.}
\label{tab:physconst}
\centering
\begin{tabular}{c c c}
\hline\hline
Quantity & Value & Unit \\
\hline
Solar mass M$_\odot$ & $1.9890 \times 10^{30}$ & kg \\
\smallskip
Solar radius R$_\odot$& $6.95508 \times 10^{8}$ & m \\
\smallskip
Solar luminosity L$_\odot$& $3.846 \times 10^{26}$ & kg~m$^2$~s$^{-3}$ \\
\smallskip
Gravitational constant $G$& $6.67232 \times 10^{-11}$ & m$^{3}$~kg$^{-1}$~s$^{-2}$ \\
\end{tabular}
\tablefoot{Solar mass and gravitational constant are defined based on the measurement of $G$M$_\odot$ \citep[see e.g.,][and references therein]{2005MNRAS.356..587C}.}
\end{table}
\section{Stellar evolution codes}\label{sec_evocodes}
Results have been computed with different stellar evolution codes that are widely used for producing sets of tracks and isochrones and have been applied in a broad range of fields of astrophysics. We use the {\ttfamily fgong} file format as the standard to export and compare interior models, which includes a comprehensive set of global stellar properties and quantities of interest for stellar evolution and asteroseismic comparisons. Tracks in {\ttfamily ASCII} format including the complete evolution are also compiled and available on the website.

The stellar evolution codes participating in the {\it Aarhus Red Giants Challenge} are: {\tt ASTEC} \citep{ChristensenDalsgaard:2008bi}, {\tt BaSTI} \citep{Pietrinferni:2004im}, {\tt CESAM2k} \citep{2008Ap&SS.316...61M}, {\tt GARSTEC} \citep{Weiss:2008jy}, {\tt LPCODE} \citep{Althaus:2003cx}, {\tt MESA} \citep{Paxton:2011jf}, {\tt MONSTAR} \citep{Constantino:2015fu} , {\tt YaPSI} \citep{Spada:2017ev}, and {\tt YREC} \citep{Demarque_ea:2008}. For the interested reader, a detailed description of the codes including additional ingredients entering the calculations (such as neutrino losses) is given in Appendix~\ref{sec_app_codes}.
\section{Solar radius calibration}\label{sec_suncal}
Convection has been treated using the mixing-length theory in all calculations. Since the exact implementation of this theory varies across evolutionary codes \citep[see e.g.,][for a discussion]{Salaris:2008fy}, we have initially set up a solar radius calibration to determine the corresponding convective efficiency to be used in all subsequent calculations. The codes were requested to match the solar radius of $6.95508\times10^{10}$~cm at an age of 4.57~Gyr by tuning the mixing-length parameter $\alpha_\mathrm{MLT}$ ($l_\mathrm{MLT}=\alpha_\mathrm{MLT} \times H_P$, where $H_P$ is the local pressure scale height) and fixing the initial hydrogen abundance to $X=0.7$. Given that microscopic diffusion is not included in the standard input physics described in Section~\ref{sec_inpphys}, the solar surface abundances $Z/X=0.0245$ from the \citet{Grevesse:1993vd} mixture define the initial composition to be $Z=0.01715$ and $Y=0.28285$. Consequently, the current solar luminosity and effective temperature were not quantities to be reproduced as part of the calibration.

The choice of this rather unusual procedure to calibrate the convective efficiency was motivated by the constraints from the adopted input physics (i.e., no microscopic diffusion), and the goal of producing stellar structures that have the same mean density as ensured by our convergence criterion in Eq.~\ref{eq:converg}. This is the reason why the solar radius was chosen as the quantity to be reproduced by tuning the mixing-length value instead of e.g., the solar luminosity.

Table~\ref{tab:suncal} presents the results of this exercise for the participating evolutionary codes. As expected, the obtained luminosity is about 20\% higher than solar and the effective temperatures hotter than the Sun by about $\sim250$~K. These discrepancies with respect to the solar values are the result of the chosen constraint and input physics, in particular the lower bulk metallicity of the models compared to that of the Sun. It is worth noticing that already at this evolutionary stage the numerical implementation of the evolutionary codes produces variations of up to $\sim40$~K and 0.03~${\rm L}_\sun$. How these differences develop at later evolutionary phases will be discussed in the following sections.
\section{Evolutionary tracks: solar-radius calibrated}\label{sec_suntracks}
\begin{figure}
\begin{center}
\includegraphics[width=\linewidth]{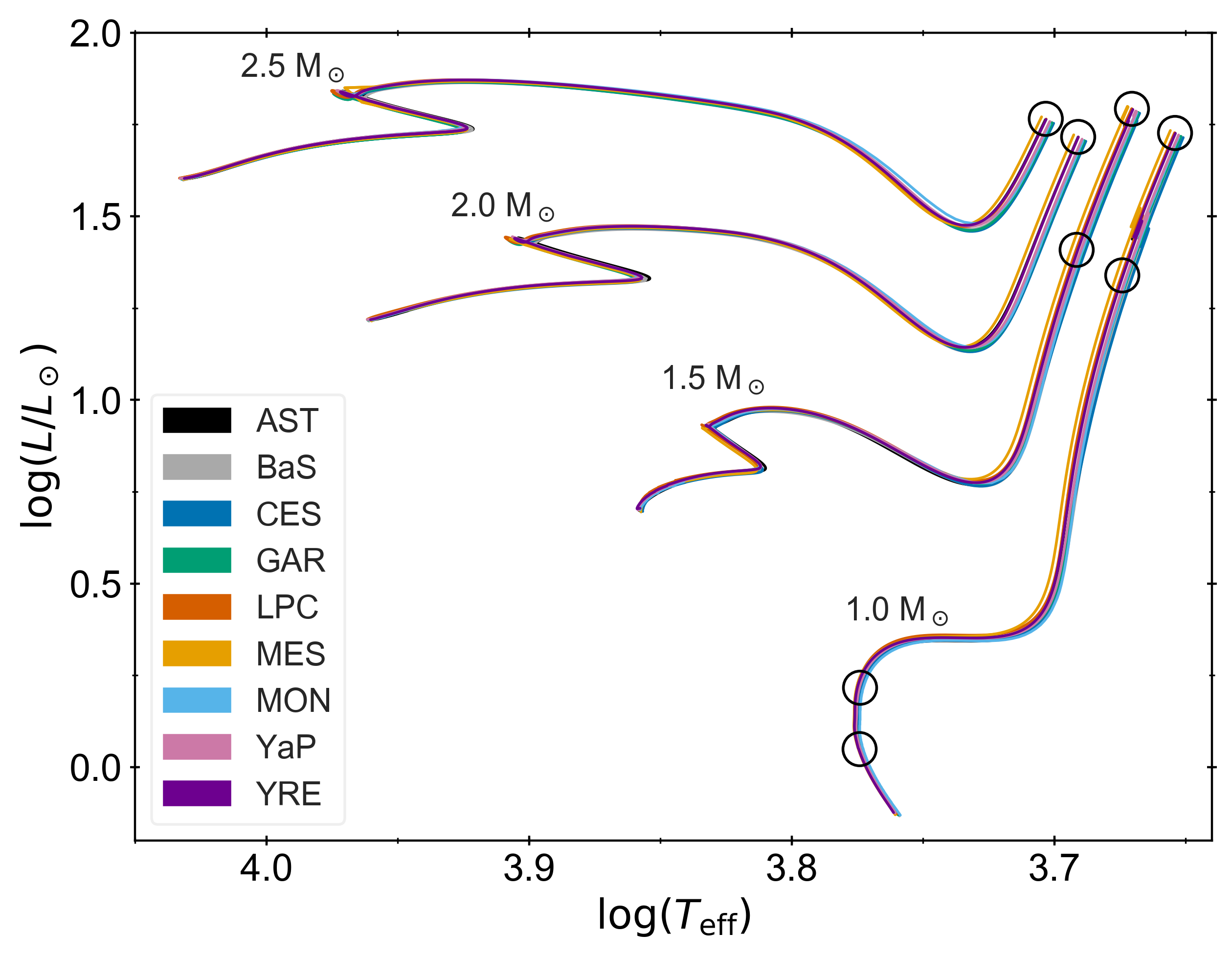}
\caption{Hertzsprung-Russell diagram (HRD) of solar-radius calibrated science cases for all participating codes. The position of the stellar models used in our comparisons are shown with open circles. See text for details.}
\label{fig:glob_suntracks}
\end{center}
\end{figure}
Using the standardized input physics, fundamental constants, and the value of the mixing-length parameter $\alpha_\mathrm{MLT}$ calibrated as described in Section~\ref{sec_suncal}, we computed our first science cases consisting of evolutionary tracks with masses 1.0,~1.5,~2.0, and 2.5~M$_\sun$ at a composition of $Y=0.28$ and $Z=0.02$. The aim of this exercise was to provide the best-case scenario of model comparisons, namely using the same input physics for all evolutionary codes after calibrating them to a common point in the main-sequence phase. {\it Differences between models will reveal the minimum level of systematic uncertainties arising solely from the numerical implementation of each evolutionary code that can be expected when determining stellar properties of red giants using e.g., isochrone fitting techniques}. We emphasize once again that this is the rationale behind choosing comparison points at fixed mass and radius. In astrophysical applications these quantities are normally not known to a high enough precision to effectively serve as benchmarks for our study.

The overall evolution of these stellar tracks is shown in Fig.~\ref{fig:glob_suntracks} where it can be seen that differences in effective temperature appear on the main sequence and further increase during the red-giant-branch phase. To better understand their origin we compared the interior structure of models at different evolutionary points (see circles in Fig.~\ref{fig:glob_suntracks} and Tables~\ref{tab:suncal10},~\ref{tab:suncal15},~and~\ref{tab:suncal2025}), which we analyse in detail in the following sections. We note that the change in composition with respect to the solar radius calibration values was motivated by the existence of original opacity tables at $Z=0.02$ and thus aiming at minimizing differences due to interpolations in the opacity calculations across tables.
\subsection{Evolution of 1.0~M$_\sun$ models}\label{ssec_sun1M}
Besides models at a fixed value of stellar radius, we also consider a comparison point at the end of the main-sequence phase (the Terminal Age Main Sequence -- TAMS), which we define the stage where the fractional hydrogen abundance in the centre reaches $X_\mathrm{c}=1\times10^{-5}$. The tracks are computed from the zero-age main sequence, but some of them include the final phases of the pre-main sequence to produce a properly converged initial model. The net result is that the age across codes in not exactly zero at the ZAMS, and we correct for this effect by defining an age of zero in all calculations at the point where $X_\mathrm{c}=0.69$.
\begin{figure}
\begin{center}
\includegraphics[width=\linewidth]{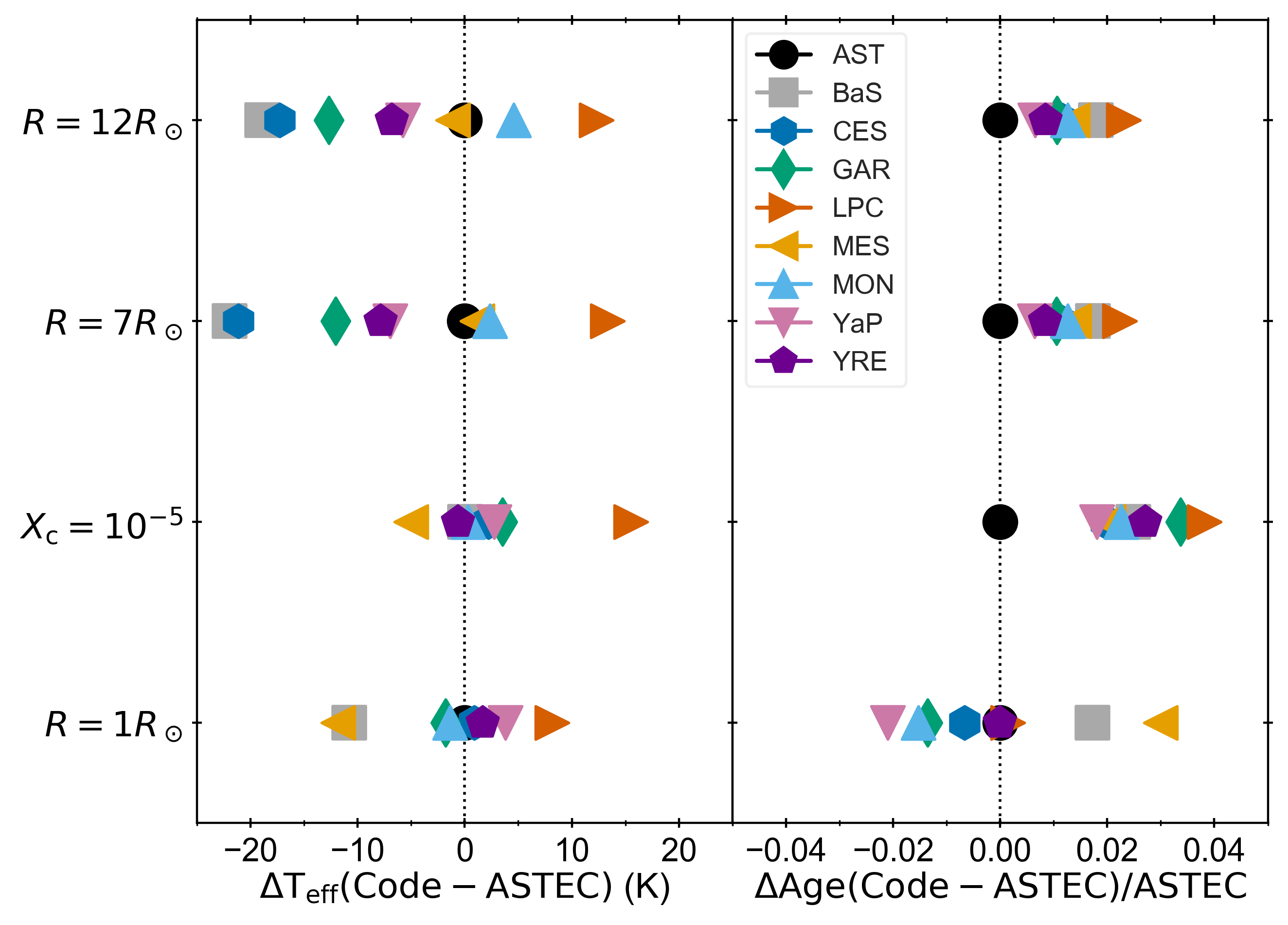}
\caption{Differences in 1.0~$\msun$ solar-radius calibrated tracks relative to {\tt ASTEC} results at different evolutionary points. {\it Left}: effective temperature differences after correcting for the calibration offset. {\it Right}: fractional age differences. See text for details.}
\label{fig:teff100suncal}
\end{center}
\end{figure}
\begin{figure*}
\begin{center}
\includegraphics[width=90mm]{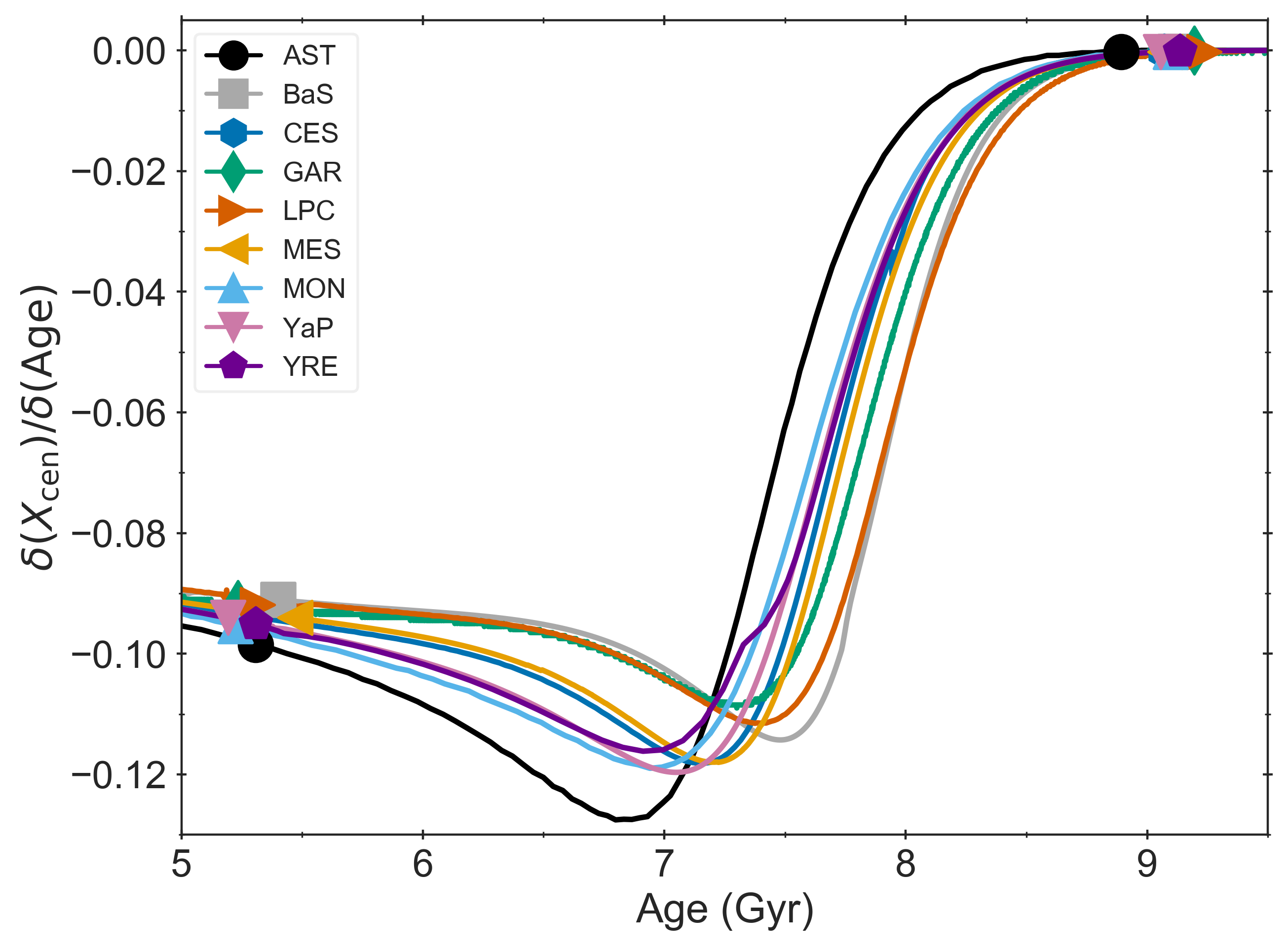}
\includegraphics[width=90mm]{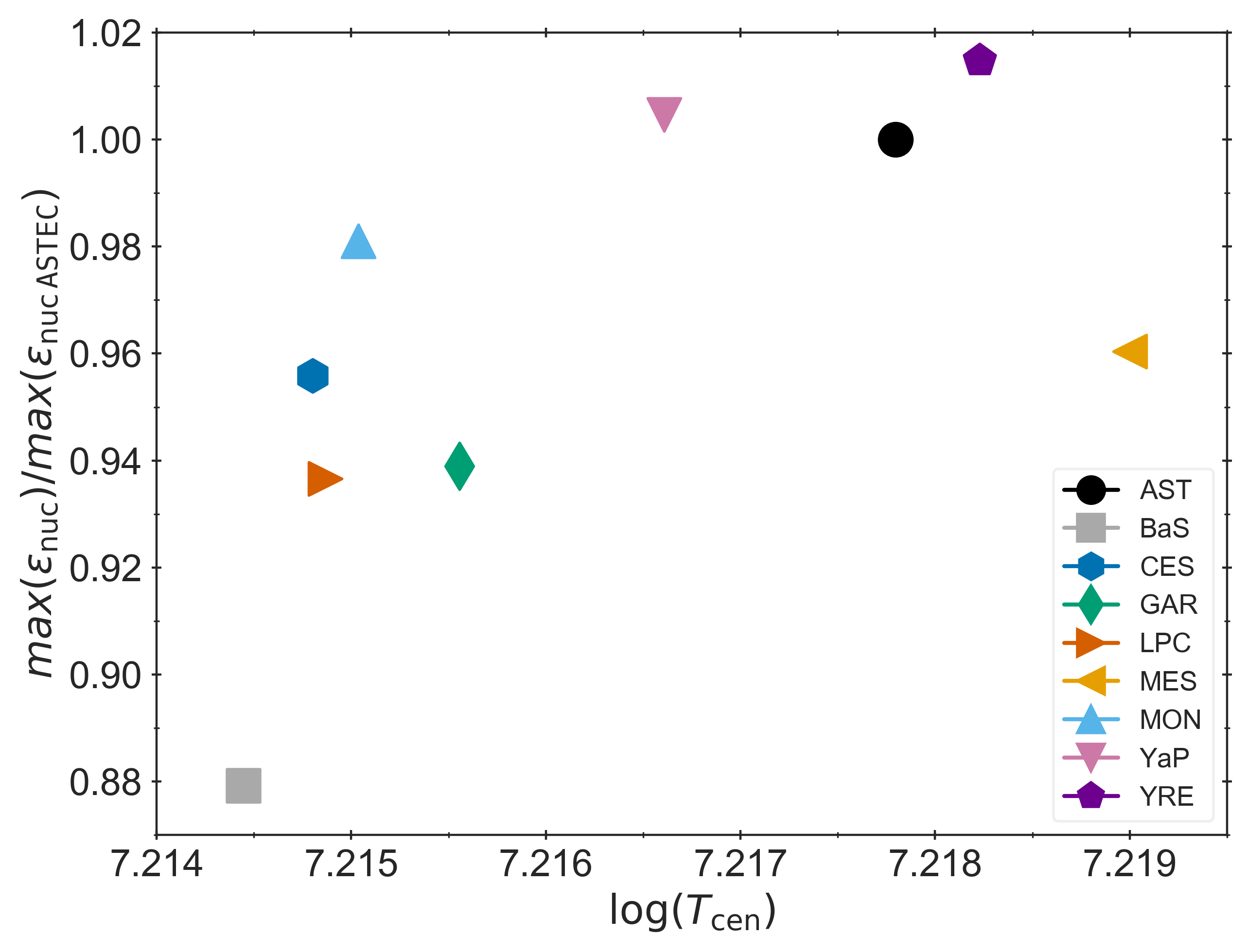}
\caption{{\it Left:} time derivative of the central hydrogen content $X_\mathrm{cen}$ as a function of age for the 1~$\msun$ solar-radius calibrated tracks. Symbols depict the location of models at 1~$\rsun$ and at the end of the main-sequence phase. {\it Right:} ratio of the maximum nuclear energy generation rate for each code relative to the {\tt ASTEC} value as a function of the central temperature for the 1~$\msun$ at 1~$\rsun$ solar-radius calibrated models.}
\label{fig:Age_dXdAge}
\end{center}
\end{figure*}

As mentioned in Section~\ref{sec_suncal}, our resulting solar radius calibration models have the desired solar radius and age but different effective temperatures. In order to quantify the temperature differences in other evolutionary phases after changing the composition, we subtract the $\teff$ difference between the solar radius calibration models of each code and that of {\tt ASTEC}, which was chosen as the reference. The left panel in Fig.~\ref{fig:teff100suncal} shows effective temperature differences at the considered evolutionary points relative to the {\tt ASTEC} results with the offset arising from the solar radius calibration already subtracted. The slight change in chemical composition in this exercise ($Y=0.28$, $Z=0.02$) compared to the one used in the solar radius calibration ($Y=0.28285$, $Z=0.01715$) already induces a divergence in the predicted effective temperature across codes at the level of $\sim$20~K for models at 1~R$_\odot$. These differences are sustained when models reach the TAMS but have increased in the RGB phase to about 40~K. We consider this 20~K and 40~K spread as the minimum level of systematic uncertainty introduced by evolutionary codes in the effective temperature scale of 1~M$_\sun$ main-sequence and RGB evolutionary tracks, respectively. This is a remarkable result showing the level of precision in effective temperature that can currently be achieved by stellar models.

Although the solar radius calibration was performed at a fixed age of 4.57~Gyr, changing the chemical composition in the science cases with respect to the one adopted in the calibration results in models reaching the solar radius after $\sim5.3$~Gyr. This is primarily due to their higher metallicity, $Z=0.02$ compared to $Z=0.01715$, leading to a higher opacity $\kappa$ and a lower stellar luminosity, and therefore a slower main-sequence evolution. At 1~R$_\sun$ age differences of $\sim$5\% are already evident, but interestingly enough once the models reach the end of the main sequence all codes show a remarkable agreement in age that is systematically older than the {\tt ASTEC} results. Once the evolution proceeds towards the red-giant branch the age scatter reaches the 2\% level and remains constant in this phase.

The reason for this variation is related to the efficiency in the conversion of hydrogen into helium during the main-sequence phase assigned by each code. The left panel of Fig.~\ref{fig:Age_dXdAge} shows the derivative of the central hydrogen abundance with respect to age as a function of age, where differences in the evolutionary speeds across codes are already visible after $\sim$5~Gyr of evolution. The {\tt ASTEC} track has a much steeper slope than the rest of the codes, which results in a quicker evolution and younger age when the TAMS is reached. The right panel of Fig.~\ref{fig:Age_dXdAge} compares the maximum value of nuclear energy generation relative to {\tt ASTEC} as a function of central temperature. The general trend is that the codes that evolve slowest ({\tt BaSTI}, {\tt GARSTEC}, and {\tt LPCODE}) present the lowest central temperature and value of nuclear energy generation, while conversely the quickest to evolve ({\tt ASTEC}) has one of the highest. This difference can be tracked down to the energy generation routines used in each code: we have recomputed, using the {\tt ASTEC} routine, the value of $\epsilon$ by adopting the thermal, density and chemical stratification originally provided by each code, and found differences with the original values of similar size as those reported in Fig.~\ref{fig:Age_dXdAge}. Nevertheless, this trend in speed of evolution cannot be solely ascribed to the energy generation routines, as for example the {\tt MESA} model evolves comparatively slow despite its high central temperature value. The remaining differences may come from e.g, the interpolation schemes used for extracting the opacity values.

In order to quantify the efficiency of energy conversion we computed the ratio between the integral of the luminosity with respect to time along the evolution until the analysed model and the amount of hydrogen processed during that period of time: 
\begin{equation}
\centering
\Lambda = \frac{\int_{0}^{T} L dt}{\mathrm{c^2} \left(\int_{0}^{M} X_\mathrm{t=0} dm - \int_{0}^{M} X_\mathrm{t=T} dm \right)}\,.
\label{eq:lumrat}
\end{equation}

All physical input being the same, the differences in luminosity should be traceable to differences in the amount of hydrogen converted to helium, and therefore the dimensionless ratio $\Lambda$ in Eq.~\ref{eq:lumrat} should be of constant value across codes. The results are given in Table~\ref{tab:suncal10} where differences up to $\sim$5\% around the median are seen in the RGB phase across all evolutionary codes in agreement with the differences found in the maximum energy generation rates.

\begin{figure}
\begin{center}
\includegraphics[width=\linewidth]{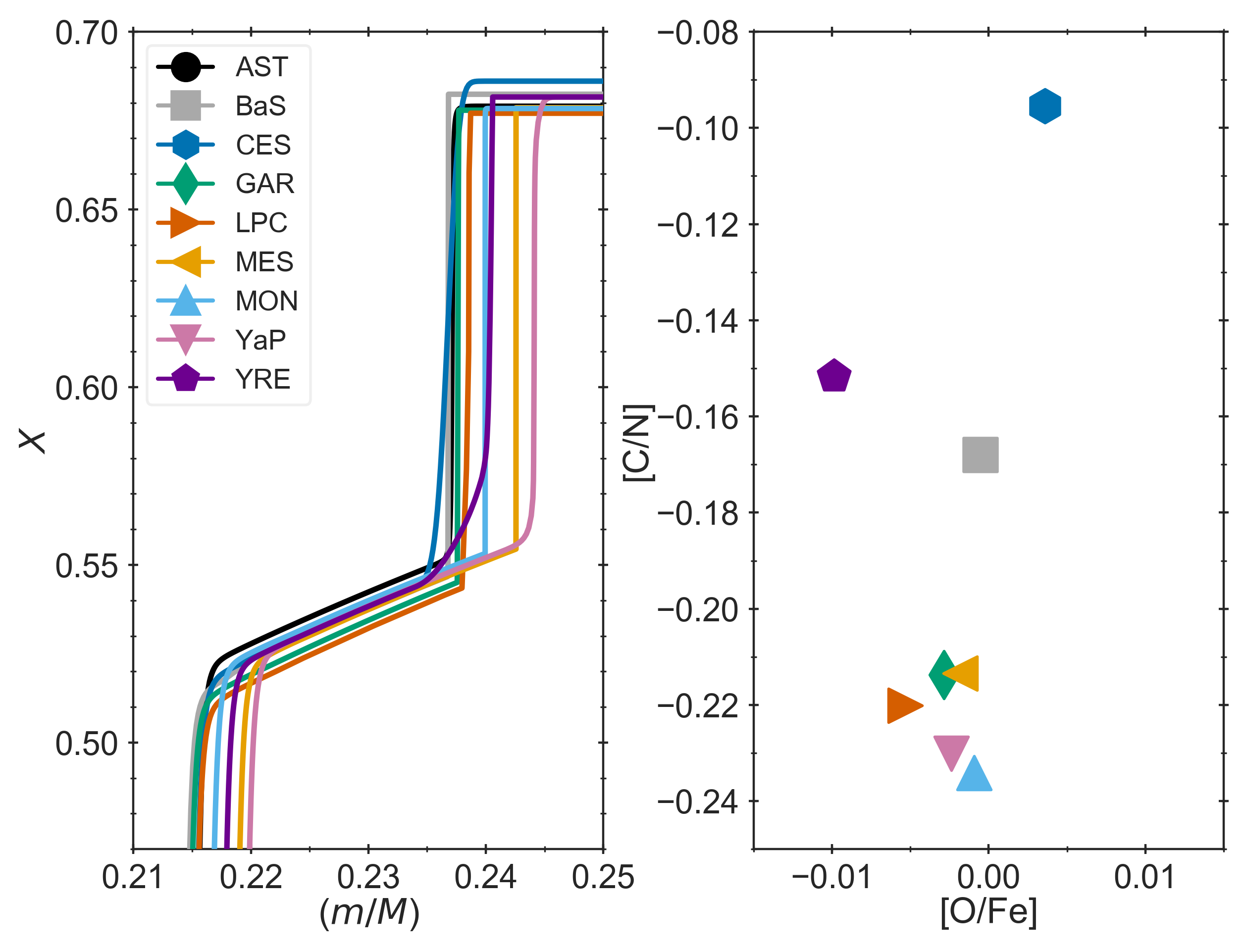}
\caption{Composition characteristics of 1~$\msun$ solar-radius calibrated models at 7~$\rsun$. {\it Left}: the interior hydrogen profile as a function of mass zoomed in around the point of deepest penetration of the convective envelope during the first dredge-up. {\it Right:} the ratio [C/N] as a function of oxygen abundance [O/Fe] at the surface.}
\label{fig:Xprof100R07suncal}
\end{center}
\end{figure}

The interior hydrogen profile in the region comprising the edge of the helium core and the point of deepest penetration during the first dredge up are shown in Fig.~\ref{fig:Xprof100R07suncal}, together with corresponding surface values of [O/Fe] and [C/N]. The latter quantity is of particular importance in the field of Galactic archaeology as it depends on the depth reached by the first dredge up, which is in turn highly sensitive to stellar mass \citep[see e.g.,][]{2015A&A...583A..87S,2016MNRAS.456.3655M}. Under these assumptions, masses (and therefore ages) of red giants can be readily extracted from observed [C/N] abundances after model-dependent calibrations of the mass-[C/N] relation as a function of metallicity \citep{2016ApJ...823..114N}. Our evolutionary codes predict a scatter of about 0.15~dex in the [C/N] ratio for the same metallicity and input physics, which comes partly from the different dredge-up depths seen in the hydrogen profiles in Fig.~\ref{fig:Xprof100R07suncal}. The scatter in the predicted [O/Fe] is below 0.02~dex and smaller than the uncertainties reported in alpha-element abundances by large spectroscopic surveys \citep[e.g., APOGEE,][]{2018ApJS..235...42A}.

\begin{figure}
\begin{center}
\includegraphics[width=\linewidth]{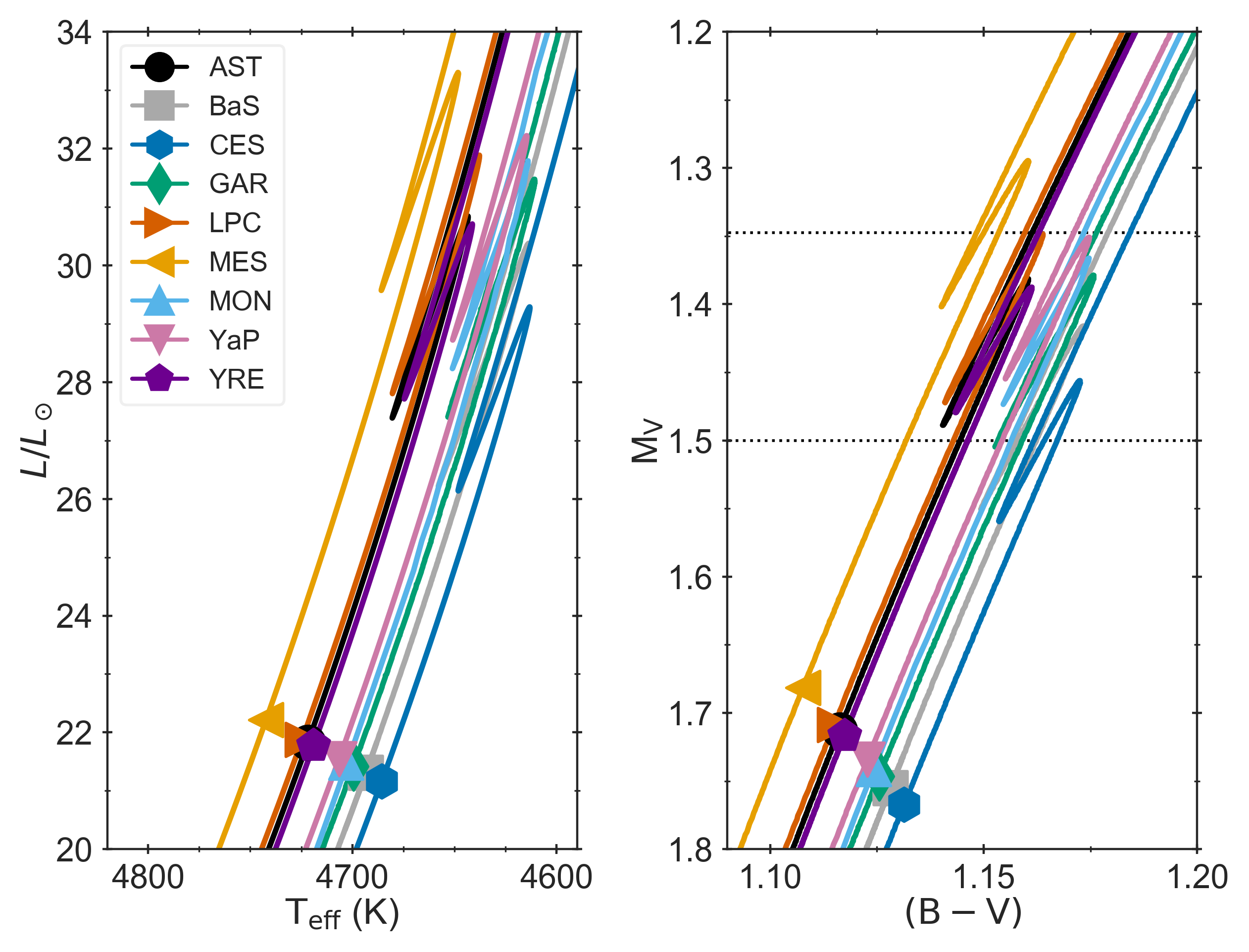}
\caption{{\it Left}: HRD depicting the 1~M$_\sun$ solar-radius calibrated tracks and the position of models at 7~R$_\sun$. {\it Right:} Optical band CMD of the 1~M$_\sun$ solar-radius calibrated tracks and corresponding models at 7~R$_\sun$. Horizontal dashed lines mark the position of the brightest and faintest predicted bump absolute magnitudes across the codes to guide the eye.}
\label{fig:Lbump100R07suncal}
\end{center}
\end{figure}

The discontinuity in composition left behind by the inwards penetration of the convective envelope during the first dredge-up is eventually reached by the advancing H-burning shell, resulting in a decrease in the luminosity known as the RGB-bump \citep[see e.g.,][and references therein]{ChristensenDalsgaard:2015bh}. The position of the RGB-bump is particularly interesting in studies of stellar clusters because isochrone fitting techniques aim at reproducing its observed luminosity. Current state-of-the-art evolutionary calculations predict bump locations $\sim$0.2~mag brighter than observed in clusters \citep[e.g.,][]{Cassisi:2011dt,Angelou:2015fk}. The results of our exercise show differences up to $\sim$4~L$_\sun$ (or $\sim$13\%) in the bump luminosity across codes, with the {\tt MESA} model reporting the highest value and {\tt CESAM2k} the lowest. We transformed the tracks to the observational plane using the routine of bolometric corrections described in Section 4.2 of \citet{Hidalgo:2018dy}, and shown in the right panel of Fig.~\ref{fig:Lbump100R07suncal}. The difference in the bolometric luminosity of the RGB bump as predicted by the various codes translates into a spread of about 0.15 mag in the Johnson V~band.

One could note that this value is of the same order of magnitude of the (typical) observed discrepancy between observations and standard model predictions of the bump luminosity. For instance, {\tt BaSTI} models appear about 0.2~mag brighter than observational data for the RGB-bump in Galactic Globular Clusters \citep[see detailed discussion in][]{2011A&A...527A..59C, cs13}. Our results show that the {\tt BaSTI} models predict one of the faintest luminosity for this feature among those shown in Fig.~\ref{fig:Lbump100R07suncal}, suggesting that the difference between the observed and predicted bump luminosity cannot be ascribed to discrepancies on how evolutionary codes manage the location of the canonical convective envelope boundary, but is the direct result of a first dredge-up event that does not penetrate deep enough in all evolutionary codes for the considered input physics. Similar results were found by \citet{2018ApJ...859..156K} who analysed a sample of red giants observed with {\it Kepler} and found that overshoot from the convective envelope was required to reproduce the brightness of the RGB bump.

The next comparison point was beyond the RGB luminosity bump at a radii of 12~R$_\sun$. In this case the interior hydrogen profiles present only one near-discontinuity corresponding to the position of the hydrogen-burning shell that is slowly moving outwards in mass as the helium core continues to grow. The surface chemical composition is the same as obtained at 7~R$_\odot$ since, in the absence of any additional mixing process, there is no change in the surface composition after the RGB-bump (see Table~\ref{tab:suncal10}). Age differences remain at the same level as they were at 7~R$_\odot$ (cf., Fig.~\ref{fig:teff100suncal}) and are sustained in the subsequent evolution towards the RGB tip.
\subsection{Evolution of 1.5~M$_\sun$, 2.0~M$_\sun$, and 2.5~M$_\sun$ models}\label{ssec_sun152025M}
\begin{figure*}
\begin{center}
\includegraphics[width=60mm]{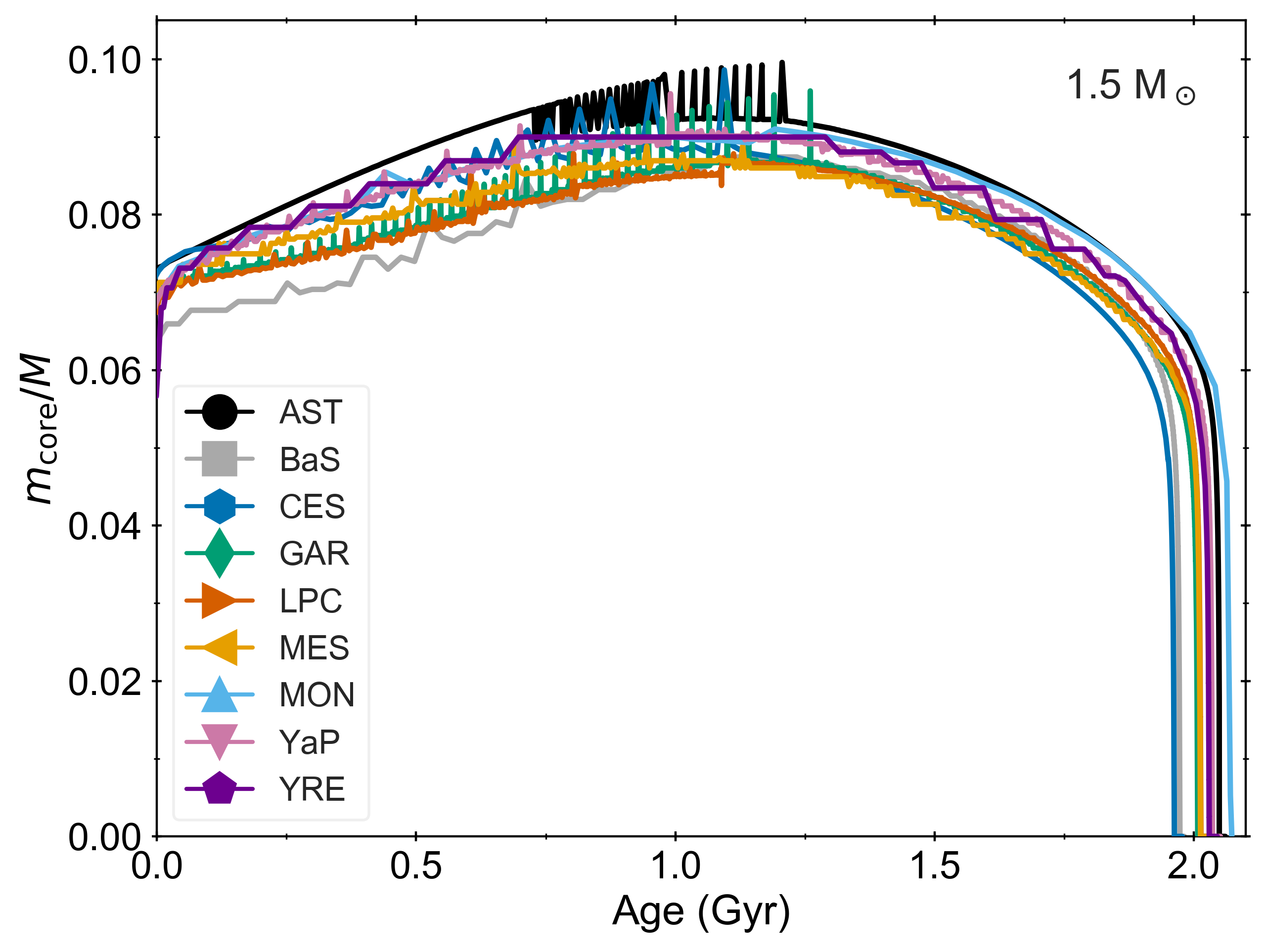}
\includegraphics[width=60mm]{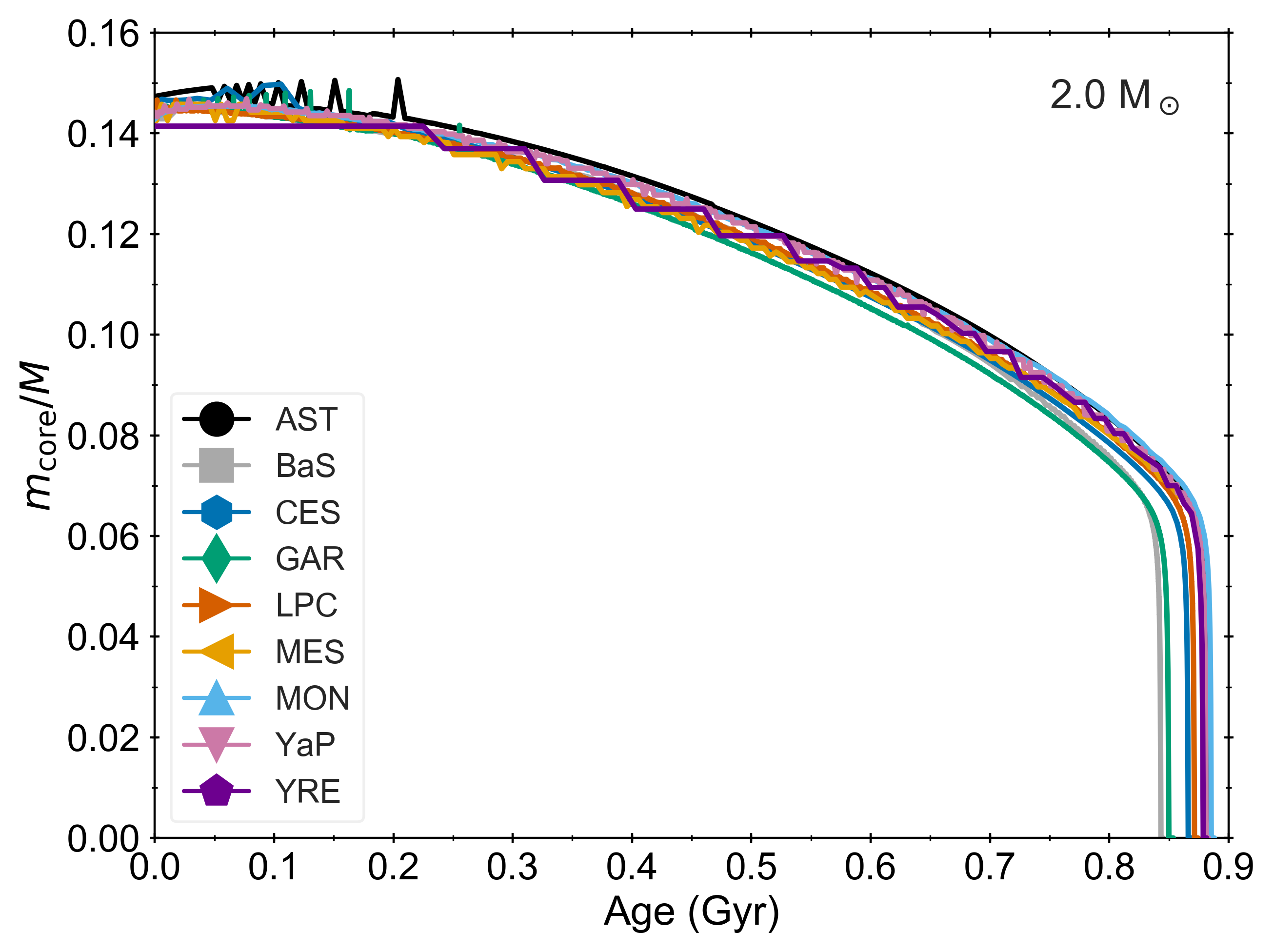}
\includegraphics[width=60mm]{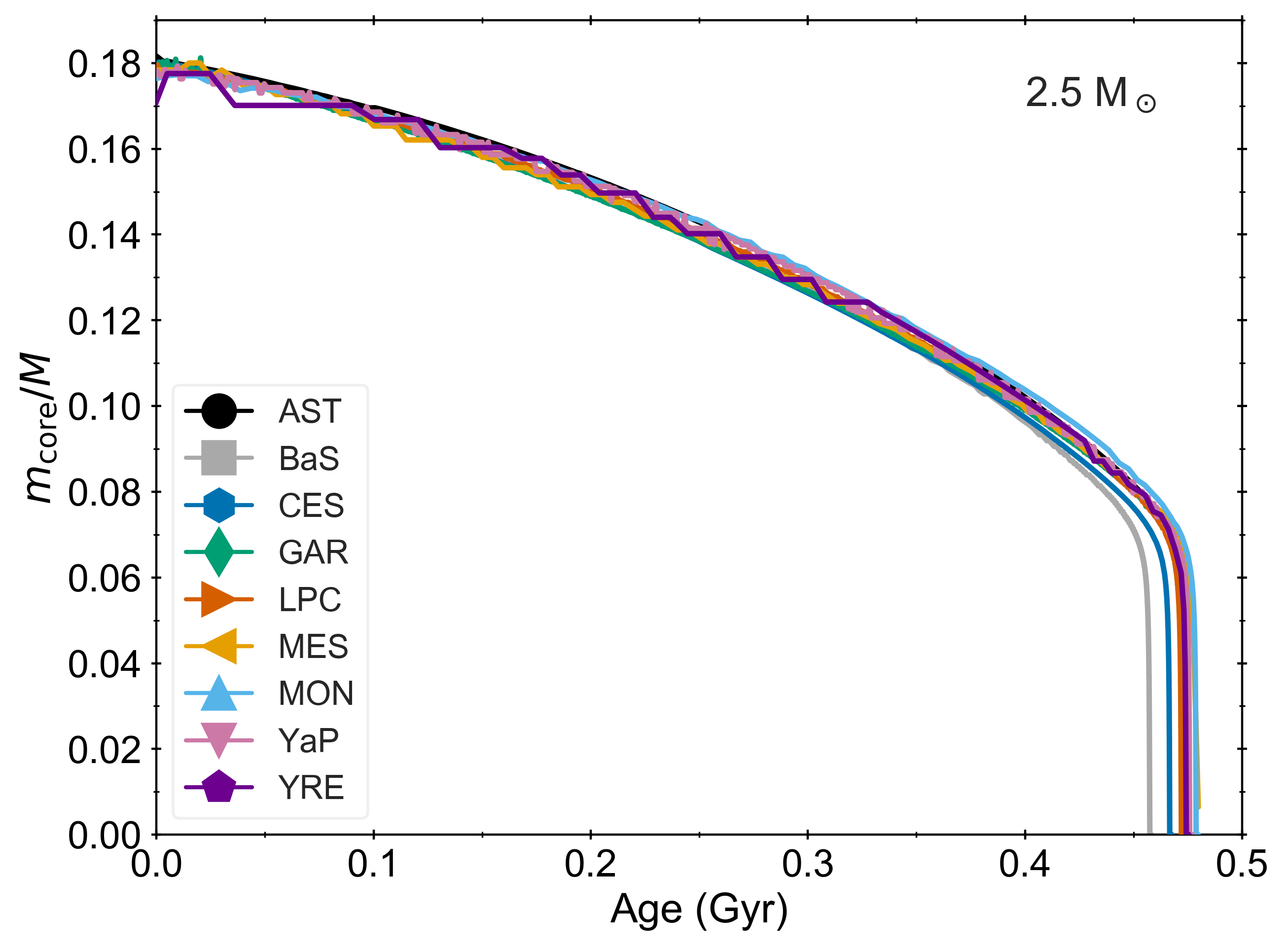}
\caption{Evolution of the convective-core size during the main-sequence phase as a function of age for the 1.5~M$_\odot$ (left), 2.0~M$_\odot$ (center), and 2.5~M$_\odot$ (right) solar-radius calibrated models.}
\label{fig:Ccoresuncal}
\end{center}
\end{figure*}

Using the same initial composition as in the 1~$\msun$ case, we calculated and compared tracks and stellar structures at higher masses: 1.5~M$_\odot$, 2.0~M$_\sun$, and 2.5~M$_\sun$. The first and most evident difference with respect to the lower-mass case is the existence of a convective core during the main-sequence evolution, whose extent as a function of age is depicted in Fig.~\ref{fig:Ccoresuncal}. Although all codes rely on the Schwarzschild criterion for defining the core convective boundary there are differences in the predicted core size of the order of 10\% across codes.

In many cases a saw-tooth profile is visible at the edge of the core as a consequence of changes in its chemical composition throughout the evolution. Nuclear burning produces changes across the chemically homogeneous central region of the model. In the case of growing convective cores (1.5~M$_\sun$), the sharp increase in density at the edge of the core increases the opacity, which consequently increases the radiative temperature gradient in that position. As evolution proceeds and the discontinuity becomes larger, the radiative gradient outside the core surpasses the value of the adiabatic gradient turning that additional layer convectively unstable. The core suddenly increases in size as is revealed by the sharp spikes in the core profiles (see left panel in Fig.~\ref{fig:Ccoresuncal}), and the subsequent fast decrease in the core extent occurs once the composition of the layer is homogenized to the core value. The radiative temperature gradient then decreases as a result of the opacity decrease and the layer becomes convectively stable again \citep[see e.g.,][]{SilvaAguirre:2011jz,Gabriel:2014bu}. The existence of these "peaks" in the core size as a function of time are the result of the numerical implementation of the Schwarzschild criterion in evolutionary codes, and the behaviour of the core boundary is expected to be smooth as a function of time in real stars. For the higher-mass cases (2~$\msun$ and 2.5~$\msun$) the evolution of the convective-core size is indeed smoother as it reaches it maximum extension at the beginning of the main sequence and contracts throughout the core hydrogen-burning phase.

One of the expected consequences of the different convective-core sizes is variations in the time spent in the main-sequence phase due to the amount of hydrogen available for nuclear burning. Models with the largest main-sequence convective cores are the oldest at the point of central hydrogen exhaustion TAMS (e.g., {\tt MONSTAR} and {\tt ASTEC} for the 1.5~$\msun$ case), while {\tt BaSTI} (and {\tt CESAM2k} for the 1.5~$\msun$ track) has the smallest core and is the youngest at the TAMS (see Fig~\ref{fig:Ccoresuncal}). Age differences at the point of central hydrogen exhaustion are of the order of 4-5\% for these masses across codes.

\begin{figure}
\begin{center}
\includegraphics[width=\linewidth]{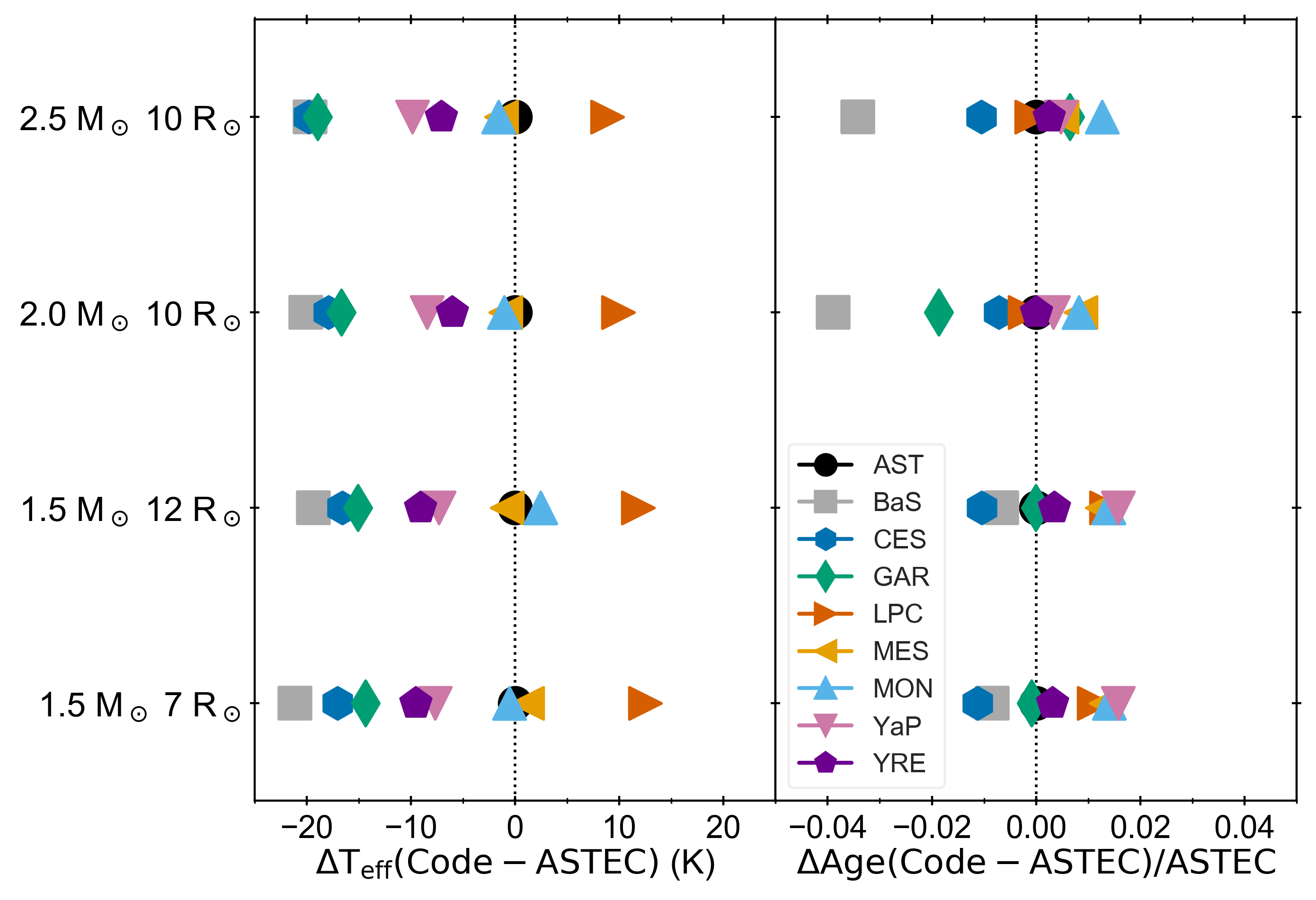}
\caption{Same as Fig~\ref{fig:teff100suncal} for 1.5~$\msun$, 2.0~$\msun$, and 2.5~$\msun$ 
solar-radius calibrated models. 
{\it Left}: effective temperature differences. {\it Right}: fractional age differences. See text for details.}
\label{fig:teff152025suncal}
\end{center}
\end{figure}

Effective temperatures and ages at the comparison points (selected radii in the RGB phase) are shown in Fig.~\ref{fig:teff152025suncal}. The variations in $T_\mathrm{eff}$ are at the $\sim$30~K level after subtracting the difference arising from the solar radius calibration model (cf., Section~\ref{ssec_sun1M}), which is similar to the results found for the 1.0~$\msun$ cases (see Fig.~\ref{fig:teff100suncal}). This reinforces the notion that differences in temperature at the RGB phase due to the numerical implementations in each code contribute to only $\sim$30-40~K regardless of the mass.

In terms of age differences, in the 1.0~$\msun$ results we found a decrease in the age scatter between the TAMS and the RGB (from $\sim$3.5\% to $\sim$2\%, see Fig.~\ref{fig:teff100suncal}). For the 1.5~$\msun$ models the decrease in age scatter goes from $\sim$5\% at the end of the main sequence (not shown in the figure) to $\sim$3\% on the RGB due to the size of the convective core: a larger mixed core in the hydrogen-burning phase translates into a longer main-sequence lifetime, which is compensated by faster evolution in the subgiant phase as the resulting helium core is closer to the Sch\"onberg-Chandrasekhar limit \citep[see][and references therein]{2015ASSP...39...11M}. This effect is still partly evident in the 2.0~$\msun$ models where the age scatter decreases from $\sim$5.3\% to $\sim$5\% between the TAMS and the RGB, and it is negligible at the higher mass end as seen by the constant age scatter of $\sim$5\% in the 2.5~$\msun$ models throughout the main-sequence and red-giant-branch evolution.

\begin{figure*}
\begin{center}
\includegraphics[width=60mm]{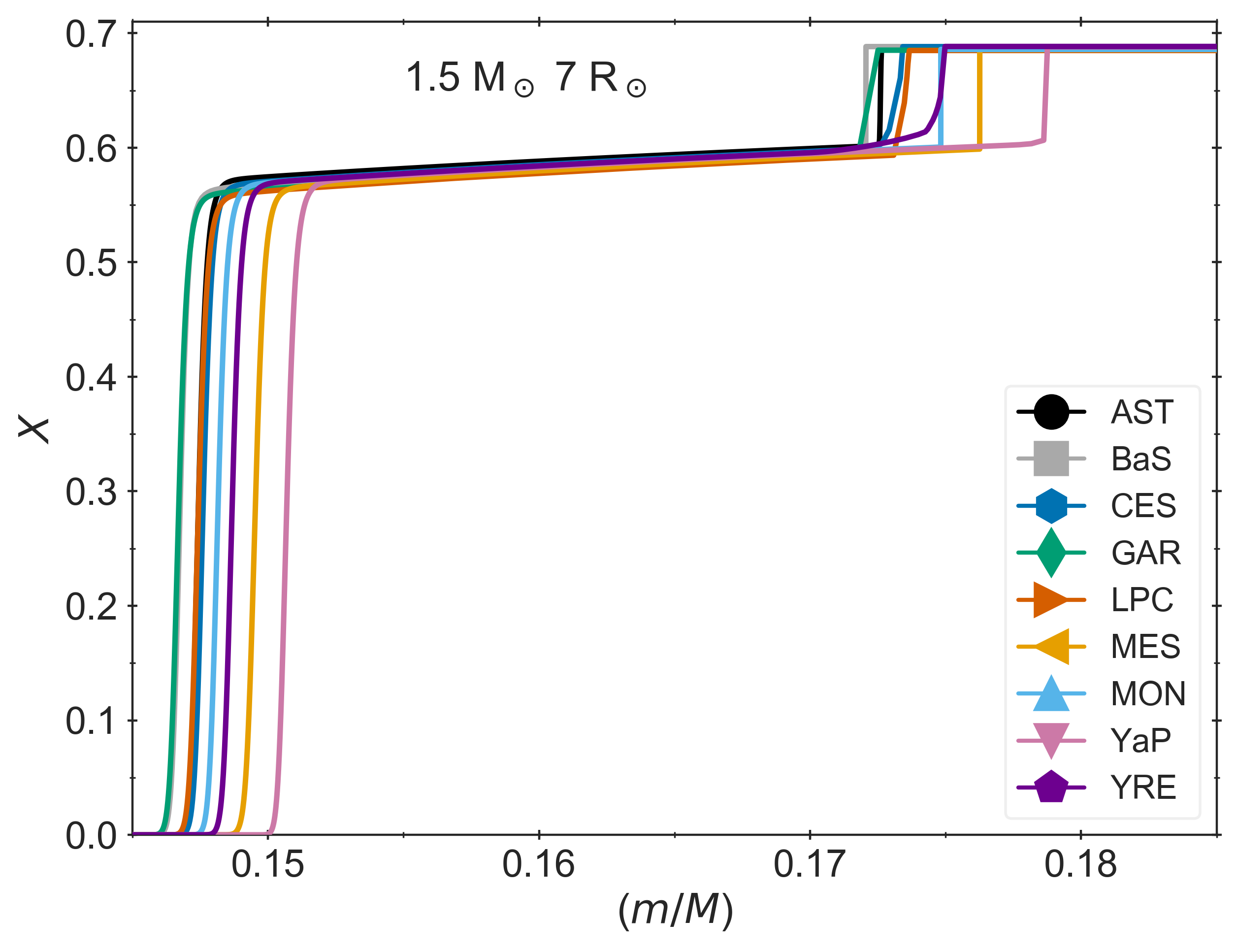}
\includegraphics[width=60mm]{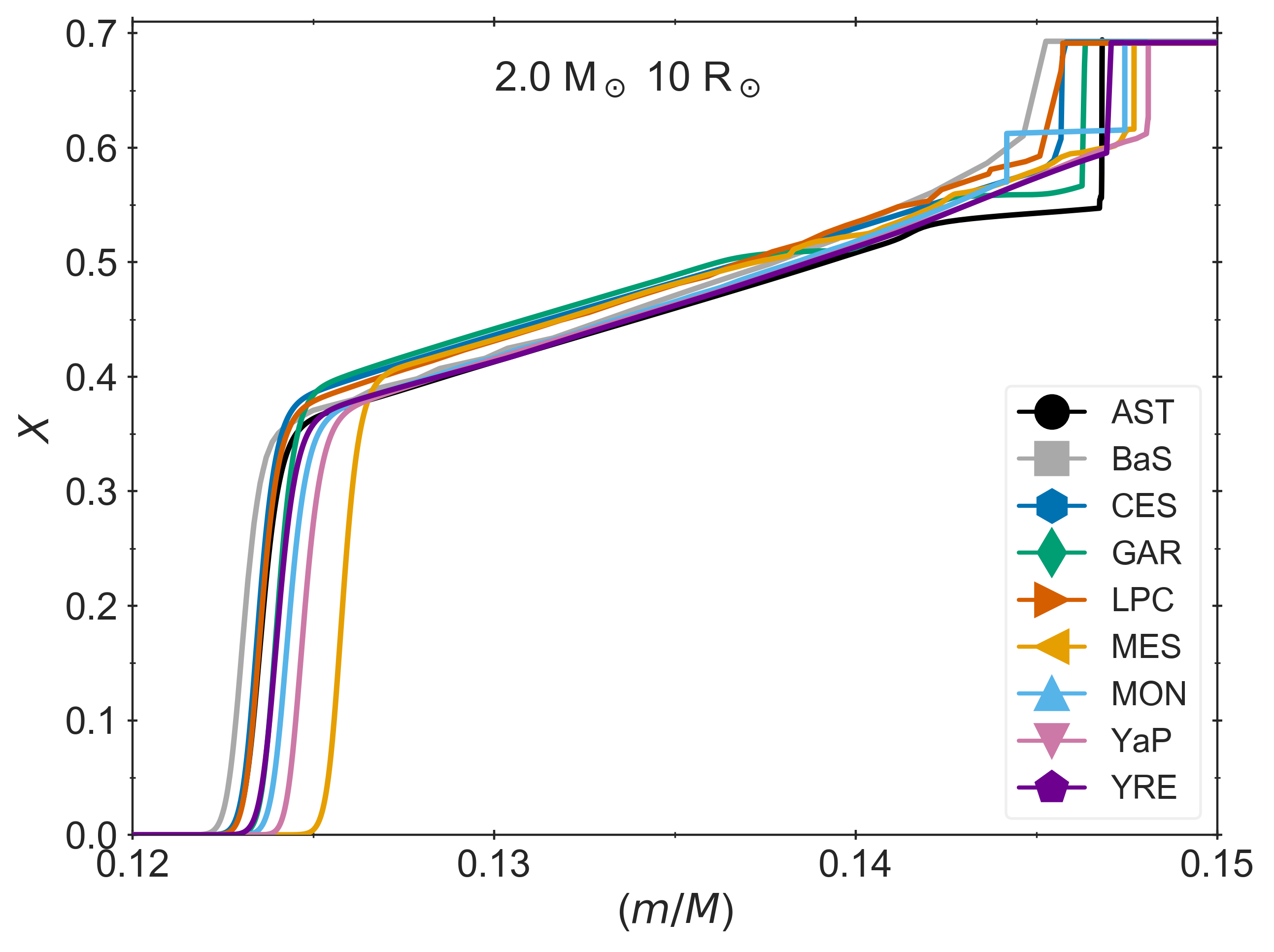}
\includegraphics[width=60mm]{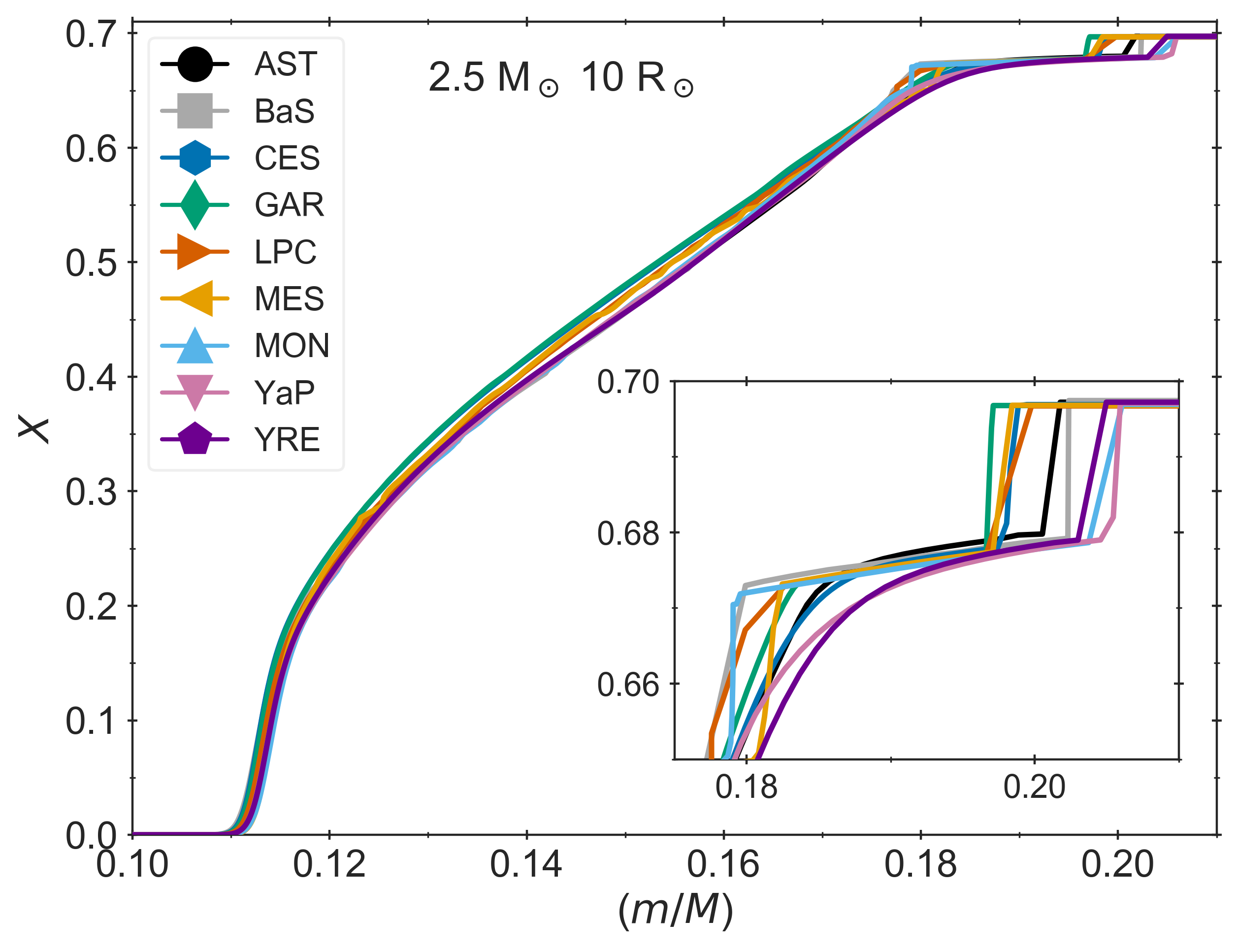}
\caption{Hydrogen profiles for models before the RGB bump of 1.5~M$_\odot$ (7~R$_\odot$, left), 2.0~M$_\odot$ (10~R$_\odot$, center), and 2.5~M$_\odot$ (10~R$_\odot$, right) solar-radius calibrated models.}
\label{fig:Xprofsuncal}
\end{center}
\end{figure*}

An interesting feature from the asteroseismology point of view is seen in Fig.~\ref{fig:Xprofsuncal}, where the interior profiles of hydrogen are shown for the following models: 1.5~$\msun$ at 7~$\rsun$, 2.0~$\msun$ at 10~$\rsun$, and 2.5~$\msun$ at 10~$\rsun$. The differences in hydrogen profiles correspond to different density distributions, which in turn translate into differences in the Brunt-V\"{a}is\"{a}l\"{a} frequency and therefore in the asteroseismic properties of the models. The point of deepest penetration of the convective envelope during the first dredge-up and the location of the H-burning shell produce a structural glitch in the Brunt-V\"{a}is\"{a}l\"{a} frequency that could be detectable from asteroseismic inference \citep{Cunha:2015im}. The impact of these differences in hydrogen profiles on the predicted frequencies of oscillation is investigated in the accompanying paper of Christensen-Dalsgaard et al. (2019).

\begin{figure}
\begin{center}
\includegraphics[width=\linewidth]{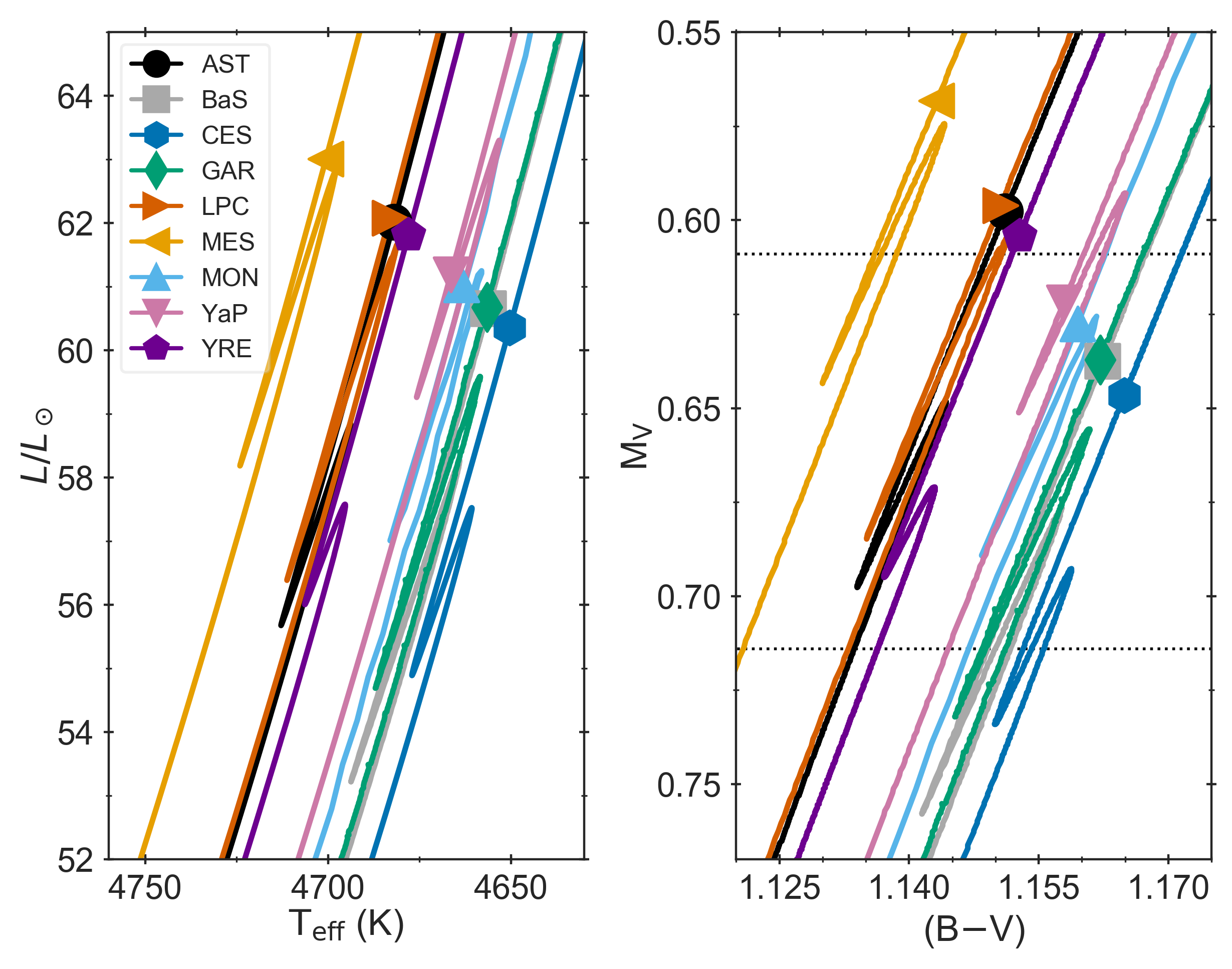}
\caption{Same as Fig.~\ref{fig:Lbump100R07suncal} for 1.5~$\msun$ tracks of solar-radius calibrated models.}
\label{fig:Lbump150suncal}
\end{center}
\end{figure}

Figure \ref{fig:Lbump150suncal} shows the position of the luminosity bump in the HRD, revealing once again differences of $\sim$6~L$_\sun$ (or $\sim$10\%) across codes that translate into $\sim$0.12~mag in the V-band. This spread is of the same magnitude as found in the 1~$\msun$ case and reinforces the notion that part of the discrepancy between observed and modelled RGB-bump luminosities is an effect of the treatment of convective boundaries in evolutionary calculations.

\begin{figure}
\begin{center}
\includegraphics[width=\linewidth]{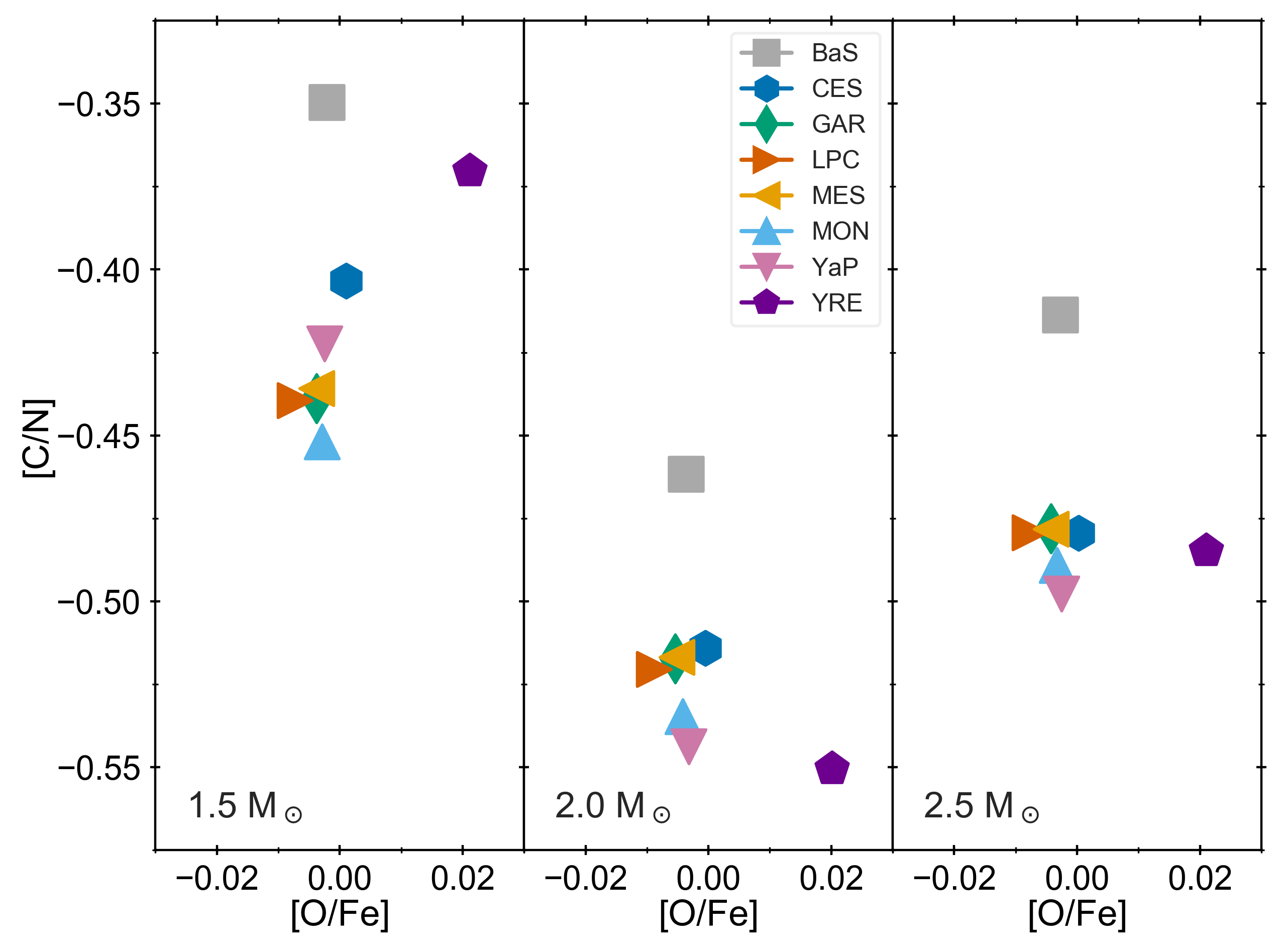}
\caption{Abundance of [C/N] as a function of oxygen abundance for solar-radius calibrated models in the RGB. From left to right: 1.5~$\msun$ at 7~$\rsun$, 2.0~$\msun$ at 10~$\rsun$, 2.5~$\msun$ at 10~$\rsun$.}
\label{fig:Chem152025suncal}
\end{center}
\end{figure}

Variations in the luminosity integral (cf., Eq.~\ref{eq:lumrat}) are of the order of 3-5\% around the median for the 1.5~$\msun$, 2.0~$\msun$, and 2.5~$\msun$ solar-radius calibrated models on the RGB. These values are at a similar level as those found in the 1~$\msun$ case, and suggest that differences are related to the implementation of the CNO-cycle during the central hydrogen-burning phase. The ratios of carbon to nitrogen as a function of oxygen for the high mass models are depicted in Fig.~\ref{fig:Chem152025suncal}. The scatter in [O/Fe] is below 0.03~dex, while the abundances of [C/N] show a spread of about 0.1~dex. This level of variation is similar to that found in the 1~$\msun$ science case and suggests that observations of the [C/N] ratio to determine masses should be treated with caution, as the calibrators needed to correlate abundances with stellar properties can suffer from systematics as those shown in these calculations.
\section{Summary and conclusions}\label{sec_summ}
We have presented the first sets of science cases for the {\it Aarhus Red Giants Challenge}, a series of workshops focused on in-depth comparison of stellar evolution and pulsation codes. We produced evolutionary tracks and structure models in the RGB phase after calibrating the mixing length parameter to the solar radius at solar age, while keeping the input physics and fundamental constants the same for all participating codes. Thus, our results can be regarded as the minimum level of systematic uncertainty in evolutionary models of red-giant-branch stars arising solely due to the numerical implementation and underlying assumptions adopted in stellar evolution codes. Our main findings can be summarized as follows:
\begin{itemize}
\item Changing the chemical composition of our models after the solar-radius calibration produces differences in effective temperature in the main-sequence phase at the 20~K level. This result suggests that theoretical temperatures are not accurate below this threshold.
\item Evolutionary tracks of different masses on the RGB present effective temperature differences of the order of 30-40~K, and age differences increasing from 2\% at 1~$\msun$ to 5\% at 2.5~$\msun$. The age scatter can be partly traced back to the energy generation routines used by each code, as also revealed by the differences in the energy produced per gram of burned hydrogen (cf., Eq.~\ref{eq:lumrat}).
\item For the 1.0~M$_\sun$ and 1.5~M$_\sun$ cases where evolution proceeds through the RGB-bump we obtain differences of the order of 10\% in the bump mean luminosity, which translates into a spread of $\sim$0.1~mag in absolute visual magnitude. However, since the bumps for the {\tt BaSTI} models, which are already brighter by about 0.2~mag than empirical measurements, are among the faintest ones, this result indicates that the first dredge-up event is not deep enough in all considered stellar evolution codes when using the input physics we adopted.
\item We find good agreement in the predicted abundance of oxygen on the RGB, but a significant spread of 0.1~dex in the [C/N] ratio. Care must be taken when using this ratio as a tracer of mass after calibration to e.g., evolutionary calculations, since there is no unique correlation between the model masses, their metallicities, and their predicted carbon and nitrogen abundances after the first dredge-up.
\end{itemize}

We would like to close this paper with a short summary about the learning process resulting from the {\it Aarhus Red Giants Challenge}. During 9 one-week workshops over the past 7 years we have been sitting together as code developers and openly compared our evolutionary codes, digging deep into routines that in some cases been have written decades ago. We have found bugs and inconsistencies in every single one of the participating codes that have helped improve our numerical and physical prescriptions as well as understanding the underlying source of differences in our results. We are certain that our codes have become more robust in this process, and we are sure that such tests and comparisons are necessary for every stellar evolution code in order to guarantee a certain level of precision. The comparisons are being extended to the helium-burning phase and will include additional input physics as part of this series, always keeping the goal in mind of providing a better representation than before of the changing properties of stars as they evolve from the main sequence to the red-giant phase.
\begin{acknowledgements}
Funding for the Stellar Astrophysics Centre is provided by The Danish National Research Foundation (Grant agreement No. DNRF106). The research was supported by the ASTERISK project (ASTERoseismic Investigations with SONG and {\it Kepler}) funded by the European Research Council (Grant agreement No. 267864). VSA acknowledges support from VILLUM FONDEN (research grant 10118) and the Independent Research Fund Denmark (Research grant 7027-00096B). AS is partially supported by grant ESP2017-82674-R (MICINN) and 2017-SGR-1131 (Generalitat Catalunya). DS acknowledges support from the Australian Research Council. Part of this research was supported by the European Research Council under the European Community’s Seventh Framework Programme (FP7/2007-2013) / ERC grant agreement no 338251 (StellarAges). RHDT acknowledges support from National Science Foundation grants ACI-1663696 and AST-1716436. This work was supported by FCT/MCTES through national funds and by FEDER - Fundo Europeu de Desenvolvimento Regional through COMPETE2020 - Programa Operacional Competitividade e Internacionaliza\c c\~ao by these grants: UID/FIS/04434/2019; PTDC/FIS-AST/30389/2017 \& POCI-01-0145-FEDER-030389. DB is supported in the form of work contract funded by national funds through Funda\c c\~ao para a Ci\^encia e Tecnologia (FCT). AM acknowledges the support of the Govt. Of India, Department of Atomic Energy, under Project No. 12-R\&D-TFR-6.04-0600. We would like to thank Bill Paxton for assistance with the MESA solar calibration and for accommodating requested changes to the code along the way. 
\end{acknowledgements}
\bibliographystyle{aa} 
\bibliography{rgwork}
%
\begin{appendix}
\section{Description of evolutionary codes}\label{sec_app_codes}
{\flushleft \tt ASTEC:}\quad the "Aarhus STellar Evolution Code" \citep{ChristensenDalsgaard:2008bi} uses an integrated treatment of the solution of the structure and chemical evolution, with time-centred differences for the dominant evolution of the hydrogen abundance, allowing fairly large time steps at adequate numerical precision. Low-temperature opacities are obtained from the \citet{Ferguson:2005gn} tables, with a gradual transition to the interior opacities around $10^4$\,K. Opacity interpolation in density and temperature uses bi-rational splines \citep{Spat91}, while interpolation in $X$ and $Z$ uses the uni-variate scheme of \citet{Akima91}. Electron screening is treated using the \citet{1954AuJPh...7..373S} formulation. Apart from the obvious inclusion in hydrogen burning, neutrino energy losses are not taken into account. In the calculation of nuclear reactions ${}^3{\rm He}$ is taken to be always in nuclear equilibrium. All initial ${}^{12}{\rm C}$ is assumed to be converted into ${}^{14}{\rm N}$ in the pre-main-sequence phase, while the gradual conversion of ${}^{16}{\rm O}$ into ${}^{14}{\rm N}$ is taken into account. The effects of nuclear burning on the overall heavy element abundance (as used in the equation of state and opacity) are not taken into account.

{\flushleft \tt BaSTI:}\quad the stellar evolution code considered for the present analysis is a slightly updated version of the one used for constructing the {\tt BaSTI} stellar models database \citep{Pietrinferni:2004im,2006ApJ...642..797P,Pietrinferni:2009dc,Pietrinferni:2013hr}, which is now capable of storing stellar structure models in the {\tt fgong} file format at any requested point during the evolution and computes the asymptotic period spacing on the fly during calculations. Concerning the input physics, the {\tt BaSTI} code treats electron screening according to the prescriptions given by \citet{1973ApJ...181..457G} and uses the \citet{Ferguson:2005gn} radiative opacities for temperatures below $10^{4}$~K. The energy losses driven by neutrinos are accounted for using the prescriptions by \citet{1994ApJ...425..222H} for the case of plasma-neutrino processes that are the dominant mechanism for neutrino emission in the cores of RGB stars; for the other mechanisms we rely on the recipes provided by \citet{1996ApJS..102..411I}. This code explicitly follows the evolution with time of the abundance of the following isotopes involved in the H-burning process: ${}^{1}{\rm H}$, ${}^{2}{\rm H}$, ${}^{3}{\rm He}$, ${}^{4}{\rm He}$, ${}^{7}{\rm Li}$, ${}^{7}{\rm Be}$, ${}^{12}{\rm C}$, ${}^{13}{\rm C}$, ${}^{14}{\rm N}$, ${}^{15}{\rm N}$, ${}^{16}{\rm O}$, ${}^{17}{\rm O}$.

{\flushleft \tt CESAM2k:}\quad the ‘‘Cesam2k stellar evolution code'' (standing for ‘‘Code d'Evolution Stellaire Adaptatif et Modulaire'', 2000 version) has been described in \cite{2008Ap&SS.316...61M}, but see also \cite{1997A&AS..124..597M} for the first version and \cite{2013A&A...549A..74M} for the recent version including the effects of rotation. The quasi-static equilibrium of a star is solved by means of a collocation method based on piece-wise polynomial approximations projected on a B-spline basis. This allows the production of the solution everywhere, not only at grid points, and also for the discontinuous variables. For the present models, the nuclear reaction network includes the reactions from the p-p chain and CNO cycle for H-burning where the abundances of ${}^{1}{\rm H}$, ${}^{3}{\rm He}$, ${}^{4}{\rm He}$, ${}^{12}{\rm C}$, ${}^{13}{\rm C}$, ${}^{14}{\rm N}$, ${}^{15}{\rm N}$, ${}^{16}{\rm O}$, ${}^{17}{\rm O}$ are followed in detail. In the convection zones mixing and evolution of chemicals are simultaneous. Energy loss due to neutrinos is accounted for following \cite{1994ApJ...425..222H} for plasma neutrinos and \cite{1966ZA.....64..395W} for photoneutrinos. Weak screening in nuclear reactions rates is accounted for following \cite{1961ApJ...134..669S}.

{\flushleft \tt GARSTEC:}\quad the "GARching STellar Evolution Code" was used for this project in a version very close to that described in \citet{Weiss:2008jy}. The main significant difference is the update of electron screening of nuclear reactions now covering the intermediate regime as well, following \citet{1973ApJ...181..439D} and \citet{1973ApJ...181..457G}. Of the different physics options described in \citet{Weiss:2008jy} we treat convective mixing as an instantaneous process. The low-temperature opacities are from \citet{Ferguson:2005gn}, while the transition and inclusion of conductive opacities is again as described in \citet{Weiss:2008jy}. The nuclear network contains for H-burning the following isotopes: ${}^{1}{\rm H}$, ${}^{3}{\rm He}$, ${}^{4}{\rm He}$, ${}^{12}{\rm C}$, ${}^{13}{\rm C}$, ${}^{14}{\rm N}$, ${}^{15}{\rm N}$, ${}^{16}{\rm O}$, ${}^{17}{\rm O}$.
 
{\flushleft \tt LPCODE:}\quad the "La Plata stellar evolution CODE" used for this work is an updated version of the one reported in  \citet{Althaus:2003cx} and \citet{MillerBertolami:2016kc} that stores complete models including the stellar atmosphere. Radiative opacities at low temperatures are from \citet{Ferguson:2005gn} and a smooth transition to the atomic OPAL opacities is carried out between $10^{4}$ and $1.25\times10^4$~K. Plasma-neutrino processes are taken from \cite{1994ApJ...425..222H} and other processes are from \citet{1996ApJS..102..411I}. Electron screening of nuclear reactions covers both the weak intermediate and strong regimes, following \citet{1973ApJ...181..457G} and \citet{1982ApJ...258..696W}. Convective mixing is always treated as a diffusive process, with the diffusion coefficient $D_c= v_c l / 3$, where $v_c$ and $l$ are the local convective velocity and mixing length respectively, obtained from the MLT. In convective regions, mixing and nuclear burning are always treated simultaneously, as described in \citet{Althaus:2003cx}. The version of {\tt LPCODE} used in the present work includes a detailed nuclear reaction network involving 32 species (including neutrons $n$ and the aluminium isomer $^{26{\rm m}}$Al) and 96 reactions for the pp chains, the CNO tricycle, the hot CNO cycle, the 3$\alpha$ and advanced $\alpha$ capture reactions, together with the most relevant neutron capture reactions. The included species are $n$, $^{1}$H, $^2$H, $^3$He, $^4$He, $^7$Li, $^7$Be, $^{12}$C , $^{13}$C, $^{14}$C, $^{13}$N, $^{14}$N, $^{15}$N, $^{16}$O, $^{17}$O, $^{18}$O, $^{19}$F, $^{20}$Ne, $^{21}$Ne, $^{22}$Ne, $^{23}$Na, $^{24}$Mg,    $^{25}$Mg, $^{26}$Mg, $^{26}$Al, $^{26{\rm m}}$Al, $^{27}$Al, $^{28}$Si, $^{29}$Si, $^{30}$Si, $^{31}$P, and $^{32}$P.

{\flushleft \tt MESA:}\quad We used version 6950 of Modules for Experiments in Stellar Astrophysics \citep[MESA][]{Paxton:2011jf,Paxton:2013km}. In addition to the physics described in Section~\ref{sec_inpphys}, we used the `extended' option for electron screening, which combines \citet{1973ApJ...181..457G} in the weak regime and \citet{1978ApJ...226.1034A} with plasma parameters from \citet{1979ApJ...234.1079I} in the strong regime. We used the low-temperature opacities from \citet{Ferguson:2005gn}, and calculations of energy loss from neutrinos are following the prescription of \citet{1996ApJS..102..411I}. The included species are $^{1}$H, $^2$H, $^3$He, $^4$He, $^7$Li, $^7$Be, $^8$B, $^{12}$C , $^{13}$C, $^{13}$N, $^{14}$N, $^{15}$N, $^{14}$O, $^{15}$O, $^{16}$O, $^{17}$O, $^{18}$O, $^{18}$F, $^{19}$F, $^{18}$Ne, $^{19}$Ne, $^{20}$Ne, $^{22}$Ne, $^{22}$Mg, and $^{24}$Mg. The {\tt inlists} used for this project are available at the workshop website.

{\flushleft \tt MONSTAR:}\quad We have implemented minor modifications to the Monash version of the Mt. Stromlo evolution code as reported in \citet{Constantino:2015fu}. Supplementary to the agreed upon input physics, stellar model calculations include the effects of weak screening \citep{1973ApJ...181..439D}. We consider the usual neutrino losses through nuclear reactions, as well as those by pair-neutrino processes, photo-neutrino processes and plasma neutrino processes \citep{1967ApJ...150..979B}, whilst bremsstrahlung rates are from \citet{1969PhRv..180.1227F}. Corrections due to neutral currents are applied according to \citet[for pair, plasma, and photo-neutrino processes]{1976MNRAS.176....9R} and \citet[for bremsstrahlung]{1976ApJ...210..481D} which impact the location of the off-center core flash. The Aesopus low-temperature opacities \citep{Marigo:2009fj} are utilised below temperatures of 10\,000~K. When computing EOS quantities and opacities near table boundaries we perform linear interpolation across datasets to ensure a smooth transition. If required, values are linearly extrapolated beyond table boundaries. For the models calculated during this workshop we instantaneously mix convective regions with mixing and burning decoupled. For the current exercise we employed a hydrogen-burning network that explicitly follows $^1$H, $^3$He, $^4$He, $^{12}$C, $^{15}$N, and $^{16}$O.

{\flushleft \tt YaPSI:}\quad These models were constructed with the same version of the Yale Rotational stellar Evolution Code (YREC) used in the Yale--Potsdam Stellar Isochrones \citep[YaPSI; see Section 2 of][]{Spada:2017ev}. The basic choices of input physics coincide with those of the YaPSI project, except for the following in order to comply with the decided choices of common physics in Section~\ref{sec_inpphys}: a) the NACRE nuclear reaction rates are adopted here; b) convective-core overshooting and microscopic diffusion are ignored; c) the Potekhin conductive opacities have been implemented; d) the \citet{Grevesse:1993vd} solar abundance mixture is adopted. The nuclear reaction network for hydrogen burning contains the following isotopes: $^1$H, $^3$He, $^4$He, $^7$Be, $^{12}$C, $^{13}$C, $^{14}$N, $^{15}$N, $^{16}$O, $^{18}$O \citep[cf.][]{Demarque_ea:2008}.

{\flushleft \tt YREC:}\quad The Yale Rotating stellar Evolution Code used to generate the models is described in \citet{Demarque_ea:2008}. In addition to the agree input physics, the {\tt YREC} version used (YREC7) uses the low-temperature opacities from \citet{Ferguson:2005gn} and includes the following species in the nuclear network: $^{1}$H, $^3$He, $^4$He, $^7$Be, $^{12}$C , $^{13}$C, $^{14}$N, $^{15}$N, $^{16}$O, and $^{18}$O. Neutrino loss rates are taken from the monograph by \citet{1989neas.book.....B}. For advanced stages of stellar evolution, the neutrino rates from photo, pair and plasma sources reported in \citet{1989ApJ...339..354I} are included.
\section{Evolutionary tracks: effective-temperature calibrated}\label{sec_tefftracks}
\begin{figure}
\includegraphics[width=\linewidth]{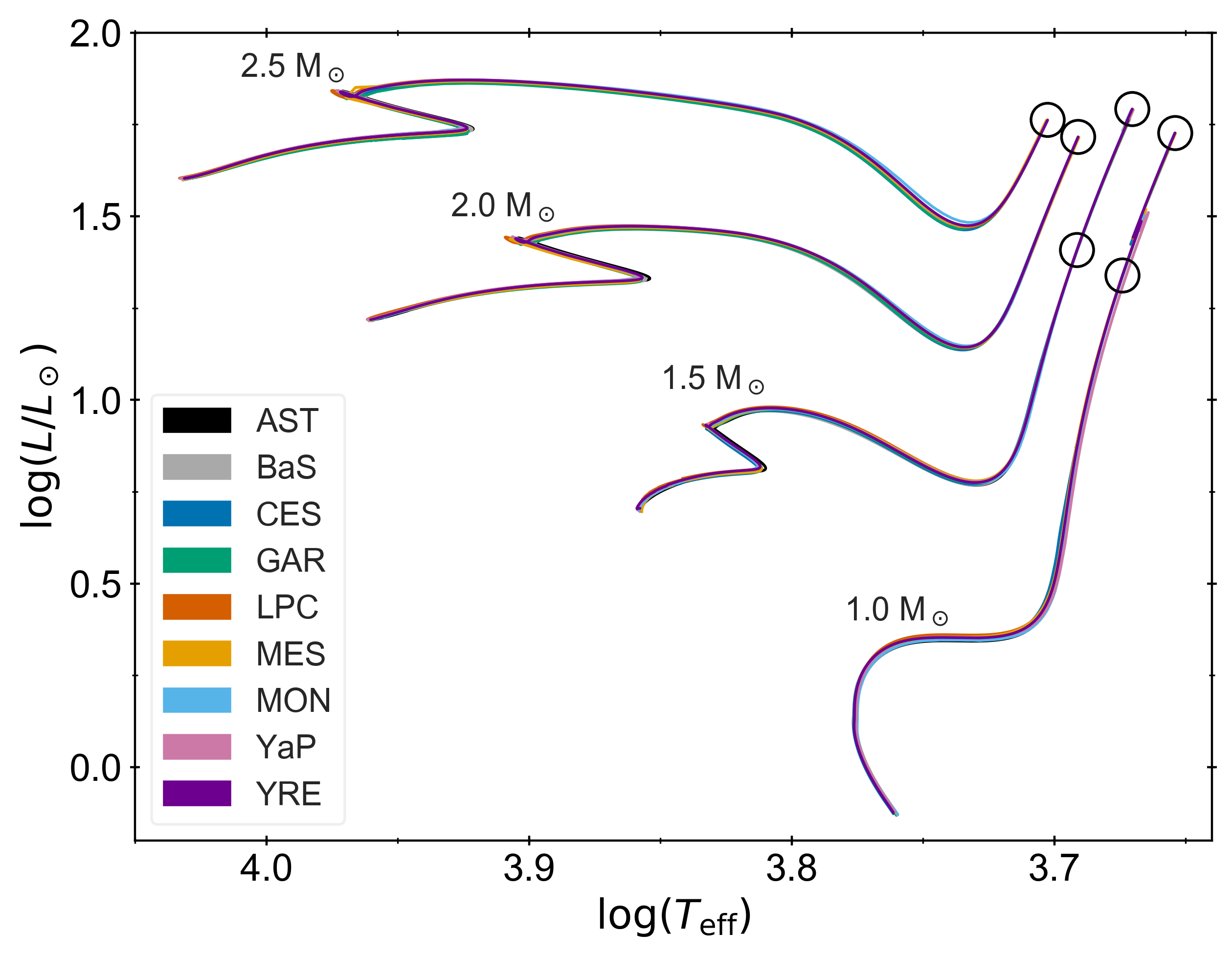}
\caption{Hertzsprung-Russell diagram of effective-temperature calibrated science cases of 1.0,~1.5, 2.0 and~2.5~M$_\sun$ for all participating codes. Open circles depict the position of models selected for detailed comparisons.}
\label{fig:glob_tefftracks}
\end{figure}
The comparisons presented in Section~\ref{sec_suncal} were made at a given mass and radius using the solar-calibrated mixing-length efficiency $\aml$. As discussed in the main text, this leads to different $\teff$ values and luminosities at a given $R$, and thus to different helium core sizes due to the core mass-luminosity relation. To uncouple the differences in the models from the details of the solar calibration, we produced an additional set of models where we calibrated $\aml$ to reproduce a certain $\teff$ value on the RGB. We chose as reference an earlier set of {\tt ASTEC} models, as they generally lie in the centre of the $\teff$ range covered by all codes as shown in Section~\ref{sec_suntracks}. The resulting HRD for the masses considered is shown in Fig.~\ref{fig:glob_tefftracks}, where we have marked the position of the selected models: 1.0~$\msun$ and 1.5~$\msun$ at 7~and~12~$\rsun$, and 2.0~$\msun$ and 2.5~$\msun$ at 10~$\rsun$.

The overall evolutionary properties in this science case are very similar to those obtained in the solar radius calibration models. However, their oscillation frequencies show interesting differences, which are investigated in the accompanying paper of Christensen-Dalsgaard et al. (2019). All models are available for the community on the website of the workshops.
\section{Main properties of calibration and science cases models}\label{sec_app_tabs}
The following tables contain the main properties of our models obtained from the modified solar calibration procedure (Table~\ref{tab:suncal}, see Section~\ref{sec_suncal}) and the solar-calibrated set of models (Tables~\ref{tab:suncal10},~\ref{tab:suncal15},~\ref{tab:suncal2025}, see Section~\ref{sec_suntracks}).
\begin{table*}[!ht]
\caption{Results of the modified solar calibration}
\label{tab:suncal}
\centering
\begin{tabular}{c c c c c c c c c c}
\hline\hline
Code & {\tt ASTEC} & {\tt BaSTI} & {\tt CESAM} &  {\tt GARSTEC} & {\tt LPCODE} & {\tt MESA} & {\tt MONSTAR} & {\tt YaPSI} & {\tt YREC} \\
\hline
\smallskip
$\alpha_\mathrm{MLT}$& 2.0437& 1.9982& 1.9496& 1.9958& 2.0431& 2.0736& 2.0489& 2.0118& 2.0388\\
\smallskip
$L/L_\sun$& 1.2032& 1.1981& 1.1913& 1.1950& 1.1947& 1.2187& 1.1864& 1.1966& 1.2073\\
\smallskip
$\teff$& 6053& 6046& 6038& 6042& 6042& 6072& 6032& 6044& 6058\\
\smallskip
$X_c$& 0.2804& 0.2874& 0.2900& 0.2903& 0.2947& 0.2880& 0.2880& 0.2909& 0.2853\\
\hline
\end{tabular}
\end{table*}

\begin{table*}
\caption{\label{tab:suncal10} Main stellar properties of 1.0~$\msun$ solar-radius calibrated models.}
\centering
\begin{tabular}{c c c c c c c c}
\hline\hline
\smallskip
$1.0~\msun$, $1~\rsun$ & $L/{\rm L}_\sun$ & $\teff$ (K)& Age (Gyr)& $m$(He)& $X_\mathrm{S}$& $Y_\mathrm{S}$& $\Lambda$ ($10^{-3}$) \\
\hline\hline
\smallskip
{\tt ASTEC}& 1.1192& 5944& 5.3077& -& 0.7000& 0.2800& 8.5756\\
\smallskip
{\tt BaSTI}& 1.1062& 5927& 5.4013& -& 0.7000& 0.2800& 8.6649\\
\smallskip
{\tt CESAM2k}& 1.1085& 5930& 5.2714& -& 0.7000& 0.2800& 8.5574\\
\smallskip
{\tt GARSTEC}& 1.1101& 5932& 5.2340& -& 0.7000& 0.2800& 8.5609\\
\smallskip
{\tt LPCODE}& 1.1173& 5942& 5.3157& -& 0.7000& 0.2800& 8.5891\\
\smallskip
{\tt MESA}& 1.1249& 5952& 5.4707& -& 0.7000& 0.2800& 8.4755\\
\smallskip
{\tt MONSTAR}& 1.1023& 5922& 5.2244& -& 0.7000& 0.2800& 8.4826\\
\smallskip
{\tt YaPSI}& 1.1158& 5940& 5.1936& -& 0.7000& 0.2800& 8.6041\\
\smallskip
{\tt YREC}& 1.1243& 5951& 5.3071& -& 0.7000& 0.2800& 8.5394\\
\hline\hline
\smallskip
$1.0~\msun$, $X_\mathrm{c}=10^{-5}$ & $L/{\rm L}_\sun$ & $\teff$ (K)& Age (Gyr)& $m$(He)& $X_\mathrm{S}$& $Y_\mathrm{S}$& $\Lambda$ ($10^{-3}$)\\
\hline\hline
\smallskip
{\tt ASTEC}& 1.6446& 5942& 8.8919& 0.0380& 0.7000& 0.2800& 7.5816\\
\smallskip
{\tt BaSTI}& 1.6783& 5935& 9.1164& 0.0350& 0.7000& 0.2800& 7.6482\\
\smallskip
{\tt CESAM2k}& 1.6715& 5929& 9.0703& 0.0376& 0.7000& 0.2800& 7.6158\\
\smallskip
{\tt GARSTEC}& 1.7313& 5935& 9.1960& 0.0394& 0.7000& 0.2800& 7.6123\\
\smallskip
{\tt LPCODE}& 1.7375& 5947& 9.2361& 0.0398& 0.7000& 0.2800& 7.6360\\
\smallskip
{\tt MESA}& 1.6733& 5956& 9.0750& 0.0371& 0.7000& 0.2800& 7.6272\\
\smallskip
{\tt MONSTAR}& 1.6694& 5921& 9.0962& 0.0405& 0.7000& 0.2800& 7.5707\\
\smallskip
{\tt YaPSI}& 1.7054& 5936& 9.0551& 0.0416& 0.7000& 0.2800& 7.6403\\
\smallskip
{\tt YREC}& 1.7128& 5946& 9.1357& 0.0419& 0.7000& 0.2800& 7.6604\\
\hline\hline
\smallskip
$1.0~\msun$, $7~\rsun$ & $L/{\rm L}_\sun$ & $\teff$ (K)& Age (Gyr)& $m$(He)& $X_\mathrm{S}$& $Y_\mathrm{S}$& $\Lambda$ ($10^{-3}$)\\
\hline\hline
\smallskip
{\tt ASTEC}& 21.829& 4721& 11.490& 0.2149& 0.6790& 0.3010& 7.2193\\
\smallskip
{\tt BaSTI}& 21.309& 4693& 11.692& 0.2140& 0.6824& 0.2975& 7.4837\\
\smallskip
{\tt CESAM2k}& 21.167& 4685& 11.619& 0.2148& 0.6861& 0.2939& 7.5385\\
\smallskip
{\tt GARSTEC}& 21.417& 4699& 11.613& 0.2142& 0.6779& 0.3020& 7.3139\\
\smallskip
{\tt LPCODE}& 21.877& 4724& 11.750& 0.2148& 0.6771& 0.3029& 7.4480\\
\smallskip
{\tt MESA}& 22.211& 4742& 11.651& 0.2183& 0.6786& 0.3013& 7.2222\\
\smallskip
{\tt MONSTAR}& 21.482& 4703& 11.637& 0.2161& 0.6965& 0.2835& 7.2142\\
\smallskip
{\tt YaPSI}& 21.548& 4706& 11.565& 0.2191& 0.6817& 0.2983& 7.3300\\
\smallskip
{\tt YREC}& 21.778& 4719& 11.587& 0.2172& 0.6816& 0.2984& 7.3741\\
\hline\hline
\smallskip
$1.0~\msun$, $12~\rsun$ & $L/{\rm L}_\sun$ & $\teff$ (K)& Age (Gyr)& $m$(He)& $X_\mathrm{S}$& $Y_\mathrm{S}$& $\Lambda$ ($10^{-3}$)\\
\hline\hline
\smallskip
{\tt ASTEC}& 53.315& 4508& 11.567& 0.2564& 0.6790& 0.3010& 7.1441\\
\smallskip
{\tt BaSTI}& 52.126& 4482& 11.776& 0.2560& 0.6824& 0.2975& 7.4326\\
\smallskip
{\tt CESAM2k}& 51.804& 4476& 11.702& 0.2571& 0.6861& 0.2939& 7.4255\\
\smallskip
{\tt GARSTEC}& 52.232& 4485& 11.692& 0.2554& 0.6779& 0.3020& 7.2736\\
\smallskip
{\tt LPCODE}& 53.392& 4510& 11.837& 0.2568& 0.6771& 0.3029& 7.4815\\
\smallskip
{\tt MESA}& 54.186& 4526& 11.725& 0.2600& 0.6786& 0.3013& 7.1583\\
\smallskip
{\tt MONSTAR}& 52.534& 4491& 11.714& 0.2578& 0.6799& 0.3000& 7.1481\\
\smallskip
{\tt YaPSI}& 52.651& 4494& 11.644& 0.2620& 0.6817& 0.2983& 7.2467\\
\smallskip
{\tt YREC}& 53.232& 4506& 11.666& 0.2595& 0.6816& 0.2984& 7.2853\\
\end{tabular}
\tablefoot{The quantity $m$(He) is the mass coordinate of the helium core, defined as the point of maximum energy release in the burning shell.}
\end{table*}

\begin{table*}
\caption{\label{tab:suncal15} Main stellar properties of 1.5~$\msun$ solar-radius calibrated models.}
\centering
\begin{tabular}{c c c c c c c c}
\hline\hline
\smallskip
$1.5~\msun$, $7~\rsun$ & $L/{\rm L}_\sun$ & $\teff$ (K)& Age (Gyr)& $m$(He)& $X_\mathrm{S}$& $Y_\mathrm{S}$& $\Lambda$ ($10^{-3}$)\\
\hline\hline
{\tt ASTEC}& 25.618& 4914& 2.5925& 0.1473& 0.6863& 0.2937& 7.2613\\
\smallskip
{\tt BaSTI}& 25.049& 4886& 2.5702& 0.1467& 0.6882& 0.2916& 7.3903\\
\smallskip
{\tt CESAM2k}& 24.953& 4882& 2.5628& 0.1475& 0.6882& 0.2917& 7.2976\\
\smallskip
{\tt GARSTEC}& 25.106& 4889& 2.5903& 0.1466& 0.6851& 0.2948& 7.2978\\
\smallskip
{\tt LPCODE}& 25.655& 4916& 2.6217& 0.1473& 0.6848& 0.2950& 7.4134\\
\smallskip
{\tt MESA}& 26.049& 4935& 2.6277& 0.1494& 0.6856& 0.2942& 7.2853\\
\smallskip
{\tt MONSTAR}& 25.165& 4893& 2.6295& 0.1481& 0.7000& 0.2800& 7.2351\\
\smallskip
{\tt YaPSI}& 25.289& 4898& 2.6341& 0.1506& 0.6880& 0.2920& 7.4037\\
\smallskip
{\tt YREC}& 25.526& 4910& 2.6007& 0.1486& 0.6883& 0.2916& 7.4358\\
\hline\hline
\smallskip
$1.5~\msun$, $12~\rsun$ & $L/{\rm L}_\sun$ & $\teff$ (K)& Age (Gyr)& $m$(He)& $X_\mathrm{S}$& $Y_\mathrm{S}$& $\Lambda$ ($10^{-3}$) \\
\hline\hline
\smallskip
{\tt ASTEC}& 62.015& 4681& 2.6493& 0.1746& 0.6863& 0.2937& 7.1827\\
\smallskip
{\tt BaSTI}& 60.657& 4656& 2.6315& 0.1741& 0.6882& 0.2917& 7.3531\\
\smallskip
{\tt CESAM2k}& 60.351& 4650& 2.6212& 0.1748& 0.6882& 0.2917& 7.2264\\
\smallskip
{\tt GARSTEC}& 60.678& 4656& 2.6492& 0.1738& 0.6851& 0.2948& 7.2601\\
\smallskip
{\tt LPCODE}& 62.075& 4683& 2.6856& 0.1749& 0.6848& 0.2950& 7.4457\\
\smallskip
{\tt MESA}& 63.006& 4700& 2.6833& 0.1770& 0.6855& 0.2943& 7.2136\\
\smallskip
{\tt MONSTAR}& 61.029& 4663& 2.6867& 0.1754& 0.6925& 0.2874& 7.1635\\
\smallskip
{\tt YaPSI}& 61.188& 4666& 2.6918& 0.1785& 0.6879& 0.2921& 7.3127\\
\smallskip
{\tt YREC}& 61.807& 4677& 2.6585& 0.1761& 0.6883& 0.2916& 7.3455\\
\end{tabular}
\end{table*}

\begin{table*}
\caption{\label{tab:suncal2025} Main stellar properties of 2.0~$\msun$ and 2.5~$\msun$ solar-radius calibrated models.}
\centering
\begin{tabular}{c c c c c c c c}
\hline\hline
\smallskip
$2.0~\msun$, $10~\rsun$ & $L/{\rm L}_\sun$ & $\teff$ (K)& Age (Gyr)& $m$(He)& $X_\mathrm{S}$& $Y_\mathrm{S}$& $\Lambda$ ($10^{-3}$)\\
\hline\hline
\smallskip
{\tt ASTEC}& 52.003& 4907& 0.9657& 0.1234& 0.6920& 0.2880& 7.2913\\
\smallskip
{\tt BaSTI}& 50.884& 4881& 0.9273& 0.1230& 0.6929& 0.2870& 7.1103\\
\smallskip
{\tt CESAM2k}& 50.627& 4875& 0.9587& 0.1234& 0.6921& 0.2877& 7.2875\\
\smallskip
{\tt GARSTEC}& 50.866& 4880& 0.9472& 0.1239& 0.6914& 0.2884& 7.1968\\
\smallskip
{\tt LPCODE}& 51.971& 4907& 0.9636& 0.1234& 0.6912& 0.2886& 7.3594\\
\smallskip
{\tt MESA}& 52.792& 4926& 0.9742& 0.1257& 0.6917& 0.2881& 7.3120\\
\smallskip
{\tt MONSTAR}& 51.066& 4886& 0.9738& 0.1242& 0.7000& 0.2800& 7.3598\\
\smallskip
{\tt YaPSI}& 51.295& 4891& 0.9690& 0.1246& 0.6919& 0.2881& 7.3454\\
\smallskip
{\tt YREC}& 51.965& 4907& 0.9658& 0.1239& 0.6916& 0.2884& 7.2701\\
\hline\hline
\smallskip
$2.5~\msun$, $10~\rsun$ & $L/{\rm L}_\sun$ & $\teff$ (K)& Age (Gyr)& $m$(He)& $X_\mathrm{S}$& $Y_\mathrm{S}$& $\Lambda$ ($10^{-3}$)\\
\hline\hline
\smallskip
{\tt ASTEC}& 58.256& 5049& 0.5001& 0.1130& 0.6972& 0.2828& 7.2159\\
\smallskip
{\tt BaSTI}& 57.061& 5023& 0.4826& 0.1121& 0.6974& 0.2824& 6.9519\\
\smallskip
{\tt CESAM2k}& 56.670& 5014& 0.4947& 0.1122& 0.6969& 0.2829& 7.1819\\
\smallskip
{\tt GARSTEC}& 56.915& 5020& 0.5034& 0.1124& 0.6968& 0.2831& 7.3280\\
\smallskip
{\tt LPCODE}& 58.174& 5047& 0.4996& 0.1126& 0.6967& 0.2831& 7.2982\\
\smallskip
{\tt MESA}& 59.095& 5067& 0.5026& 0.1135& 0.6968& 0.2830& 7.2484\\
\smallskip
{\tt MONSTAR}& 57.214& 5026& 0.5066& 0.1135& 0.7000& 0.2800& 7.3196\\
\smallskip
{\tt YaPSI}& 57.425& 5031& 0.5025& 0.1131& 0.6971& 0.2829& 7.2790\\
\smallskip
{\tt YREC}& 58.164& 5047& 0.5014& 0.1131& 0.6972& 0.2828& 7.2087\\
\end{tabular}
\end{table*}

\end{appendix}

\end{document}